\documentclass[preprint,12pt,sort&compress]{elsarticle}
\usepackage{color}

\usepackage{amssymb}
\usepackage{amsmath}
\usepackage{bm}
\usepackage{doi}
\usepackage{multirow}

\usepackage{subcaption}
\usepackage{caption}

\newcommand\figref[1]{\textbf{Fig.~\ref{fig:#1}}}
\newcommand\tabref[1]{\textbf{Table~\ref{tab:#1}}}

\journal{International Journal of Heat and Mass Transfer}

\begin{document}

\begin{frontmatter}



\title{Multi-scale topology optimization of porous heat sinks with voided lattice structure using a two-layer Darcy-Forchheimer model}


\author[label1]{Tatsuki Saito} 
\author[label2]{Yuto Kikuchi} 
\author[label1]{Kuniharu Ushijima\corref{cor1}} 
\ead{kuniharu@rs.tus.ac.jp}
\cortext[cor1]{Corresponding author}
\author[label2]{Kentaro Yaji} 

\affiliation[label1]{organization={Department of Mechanical Engineering, Tokyo University of Science},
            addressline={6-3-1,Niijuku}, 
            city={Katsushika-ku},
            postcode={1258585}, 
            state={Tokyo},
            country={Japan}}

\affiliation[label2]{organization={Department of Mechanical Engineering, The University of Osaka},
            addressline={2-1,Yamadaoka}, 
            city={Suita},
            postcode={5650871}, 
            state={Osaka},
            country={Japan}}

\begin{abstract}
This study presents a topology optimization framework for the design of water-cooled heat sinks that incorporate voided lattice structures, formulated using a two-layer Darcy-Forchheimer model. 
Conventional porous heat sinks often suffer from excessive pressure drops due to their intricate geometries, which limit their practical applicability. 
To overcome this issue, the proposed method introduces an explicit representation of both void and porous regions, together with graded lattice density, within a multi-material optimization framework. 
The two-layer Darcy-Forchheimer model enables efficient reduced-order simulations, allowing direct consideration of the heterogeneous porous-void distribution during the optimization process. 
The optimized designs are reconstructed into full-scale lattice geometries and validated through coupled thermo-fluid finite element analyses under fixed pressure-drop conditions. 
The results demonstrate that the voided lattice configurations significantly outperform conventional plate-fin and uniform lattice heat sinks, achieving approximately 20--30\% higher maximum Nusselt numbers while maintaining lower pressure losses.
%
\end{abstract}





\end{frontmatter}



\section{Introduction}
\label{sec1}
Thermal management has emerged as a critical challenge in various industrial applications, particularly in preventing the overheating of CPUs, GPUs, and lithium-ion batteries. 
Extensive research has therefore focused on the development of cooling devices such as heat sinks and heat spreaders~\cite{Heatsinkreview,Heatsinkoptreview,Carbonheatspreader}.
%
However, the ongoing demand for device miniaturization limits the available surface area for heat sinks, thereby constraining their heat dissipation capacity. 
To overcome this limitation, porous heat sinks have attracted considerable interest~\cite{doublepipeHX,FMFtopflowexp,FMFtopflow_Hongkong,Lotusheatsink}.

Porous structures, characterized by internal voids, provide a large specific surface area to enhance convective heat transfer. 
This feature makes them promising candidates for compact thermal management systems. 
Numerous theoretical, numerical, and experimental studies have investigated heat transfer in porous media~\cite{convinporousmedia,doublepipeHX,Tortuosity,FMFtopflowexp,FMFtopflow_Hongkong,Lotusheatsink}. 
%
Advances in computer-aided design (CAD) and additive manufacturing (AM) technologies have further enabled the realization of highly complex geometries. 
Among these, additively manufactured porous structures such as lattice structures have received particular attention because they combine low weight with high mechanical strength~\cite{AMreview,Ushijimalattice,Kosakalatice}.
%
Composed of interconnected beams or surfaces, such structures offer superior stiffness and vibration resistance~\cite{AMreview,Ushijimalattice,Kosakalatice}. 
More recently, their potential applications as open-cell porous media for heat sinks and heat exchangers have been widely recognized \cite{Takarazawa,B4senko,Microarchitected}.
%
Despite these advantages, porous media based heat sinks face significant challenges. 
Their intricate geometries often induce high flow resistance and pressure losses \cite{FMFtopflowexp,FMFtopflow_Hongkong,Takarazawa,B4senko}. In addition, the structural tortuosity inherent to porous networks can reduce fin efficiency \cite{Tortuosity}. 
On the other hand, recent studies have demonstrated that optimizing multi-scale structure of porous media enhances its macroscopic performance~\cite{S.Yun,B.Vaisser,S.Wang,Y.Liu,D.Li,C.Liu,LBM_naturalconv,Takezawa_lattice,Multi-scale-review}. 

%

Topology optimization, originally proposed by Bends\o e and Kikuchi~\cite{1stTO}, is now regarded as one of the most powerful design methodologies of the structure optimization. 
Borrvall and Petterson. subsequently extended its application to fluid flow problems~\cite{1stfluidTO}. 
Since then, topology optimization has been applied to a wide range of thermal-fluid systems, including both active~\cite{Yanetal,Yaji_levelset,PlateareaTO,Yaji_multifidelity} and passive~\cite{NaturalconvTO,LBM_naturalconv,NabeLKS} cooling applications. 
%

Multi-scale topology optimization of additively manufactured porous structures has recently emerged as a promising trend, enabling superior macroscopic performance \cite{Multi-scale-review}. 
In the thermal-fluid field, Takezawa et al. proposed a notable extension by employing the Brinkman-Forchheimer model~\cite{Takezawa_lattice,Takezawa_plate}, which describes fluid flow in porous media, to continuously represent the relative density of lattice unit cells. 
In their approach, the macroscopic properties of lattices with varying beam diameters are approximated as porous properties, thereby yielding an effective medium that behaves as an intermediate between solid and fluid~\cite{Takezawa_lattice,Takezawa_plate}.
However, this method assigns the porous phase across the entire design domain and therefore does not allow the existence of voids, a key feature of conventional topology-optimized heat sinks. 
The absence of explicit void regions restricts pressure-drop reduction, limiting the achievable performance gains. 
To overcome this limitation, voids must be incorporated directly within the optimization framework. 
Although a few studies have attempted to optimize both voids and non-uniform porous phases simultaneously~\cite{Fromyaji,Ozguc}, they directly treat porosity as a design variable and convert highly porous regions into voids only through post-processing. %
The post-processing may exert a significant impact on performance and hinder the generation of an optimized distribution. 

The present study addresses this issue by employing a multi-material topology optimization (MMTO) framework \cite{MMTO_enlargedSIMP,MMTO_DMO,MMTO_for_Heatconduction}, which enables the simultaneous optimization of multiple phases. 
While MMTO has previously been applied to heat sink design \cite{Nicollaetal}, the porous phase in that work is modeled as a homogeneous foamed metal, and the evaluation relied solely on macroscopic properties under the assumption of local thermal equilibrium. 

Compared to previous studies, present study utilizes MMTO framework to optimize both the distribution of void and that of graded lattice simultaneously, 
resulting in heterogeneous lattice structures containing void regions. 
Furthermore, the obtained full-scale geometries which composed of void and heterogeneous lattice structures are directly modeled, and their performance are evaluated using coupled thermal-fluid finite element simulations.
%

The remainder of this paper is organized as follows. 
Section~2 presents the formulation of the proposed method and describes each component of the framework. 
Section~3 reports and discusses the numerical results, and Section~4 summarizes the conclusions and outlines directions for future research.

\section{Formulation}
\label{sec2}

\subsection{Methodology outline}
\label{subsec2.1}

\begin{figure}[htbp]
    \centering
    \includegraphics[width=\linewidth]{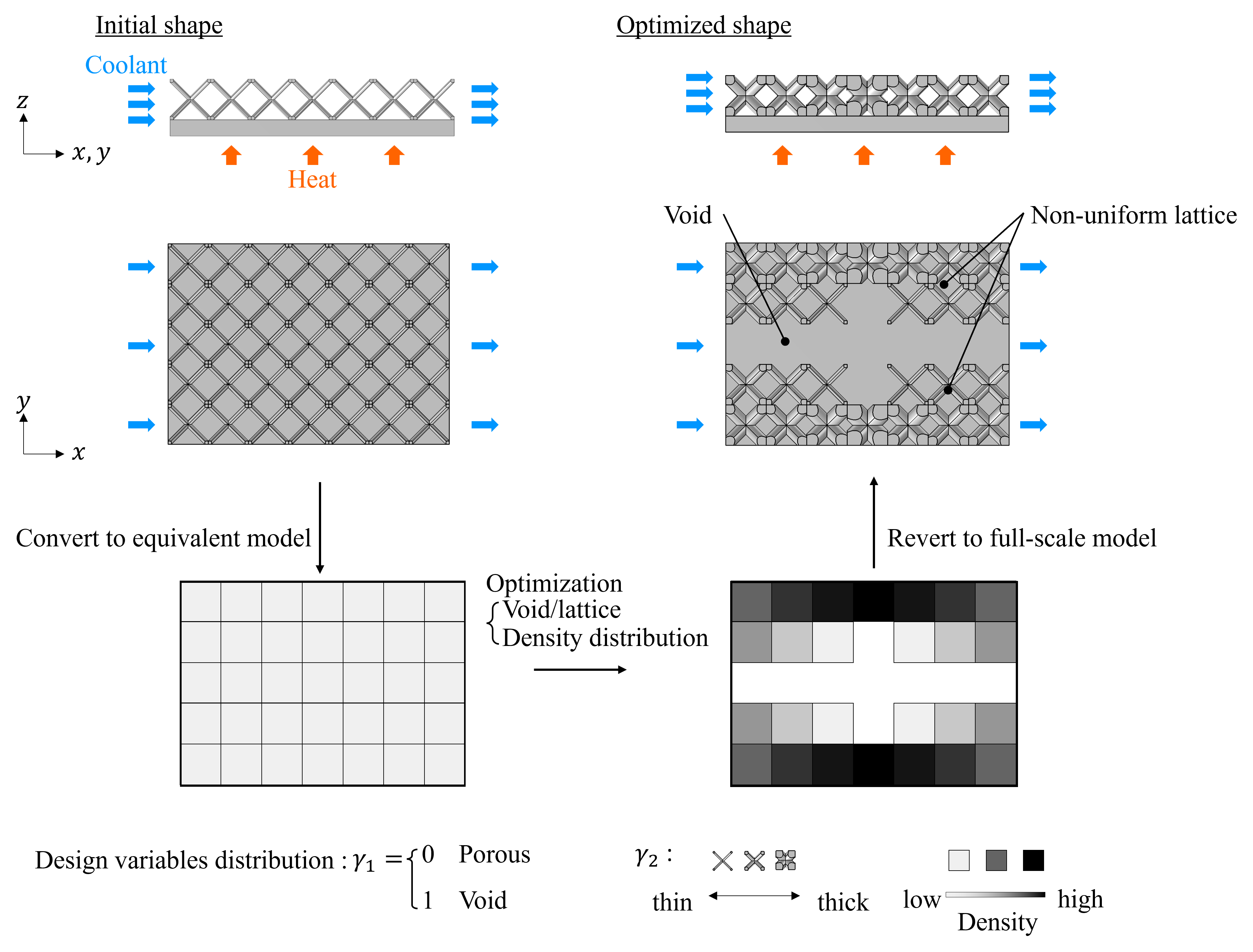}
    \caption{Schematic of the multi-scale topology optimization applied in present study. }
    \label{fig:overview}
  \end{figure}

In this section, we present the proposed optimization methods. 
An overall framework of the optimization method used in present study is illustrated in \figref{overview}. 
The objective of this work is to simultaneously optimize both the occurrence of lattice structures and the distribution of lattice density 
by using the MMTO framework. 
Two kind of design variables $\gamma_1$ and $\gamma_2$ are introduced.
The first variable $\gamma_1$ distinguishes between phases: voids are assigned where $\gamma_1 =1$,  and lattice structures are assigned where $\gamma_1 =0$. 
The second variable $\gamma_2$ represents normalized lattice 
diameter, defined as: 
  \begin{equation}
    \gamma_2=\frac{d-d_\mathrm{min}}{d_\mathrm{max}-d_\mathrm{min}}
    \label{eq:gamma2}
  \end{equation}
where $d_\mathrm{min}$ and $d_\mathrm{max}$ refer to minimum and maximum lattice beam diameters, respectively. 
Thus, denser unit cells can be arranged in larger $\gamma_2$ regions. 
The detailed methodology is presented in the subsequent subsections.
%


\subsection{Two-layer Darcy-Forchheimer model}
\label{subsec2.2}

\begin{figure}[htbp]
    \centering
    \includegraphics[width=\linewidth]{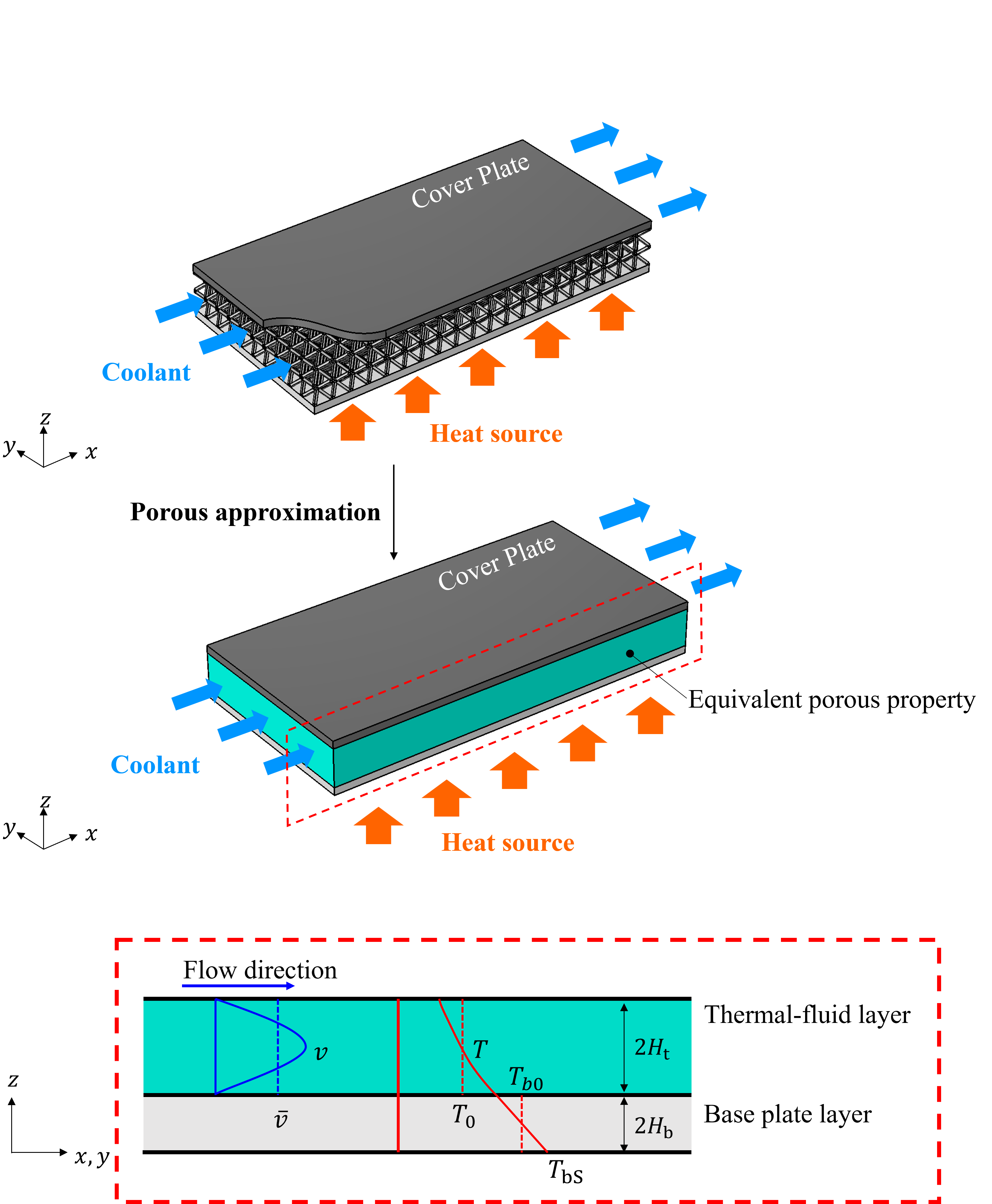}
    \caption{Overview of two-layer theory and porous approximation used in two-layer DF model. }
    \label{fig:2layerDFmodel}
  \end{figure}

Direct numerical simulations of geometrically complex non-uniform lattice media are prohibitively expensive and impractical over hundreds of optimization iterations. 
In this study, a reduced-order porous approximation, hereafter referred to as the two-layer Darcy-Forchheimer (DF) model, is employed to address this issue. 
This model builds on the theoretical framework of two-layer model proposed by Yan et al. \cite{Yanetal} and is extended in the present study by incorporating porous approximation method~\cite{Takezawa_lattice}. 
The formulation assumes that the design domain is significantly larger in the in plane ($x$-$y$) directions relative to its thickness ($z$--direction). 
As shown in \figref{2layerDFmodel}, the velocity and temperature profiles of a microchannel cross-section are modeled analytically and expressed by Eqs.~\eqref{eq:velocity_field} and~\eqref{eq:temperature_field} as:  
  \begin{equation}
    \bm{v} = \frac{3}{2} \, \bar{\bm{v}}(x, y) \left[ 1 - \left( \frac{z}{H_\mathrm{t}} \right)^2 \right]
    \label{eq:velocity_field}
  \end{equation}
  \begin{equation}
    \frac{T_\mathrm{w}(x, y)-T(x, y, z)}{T_\mathrm{w}(x, y)-T_0(x, y)}=\frac{35}{416} \left( 13+\frac{8z}{H_\mathrm{t}} -\frac{6z^2}{H_\mathrm{t}^2}+\frac{z^4}{H_\mathrm{t}^4} \right)
    \label{eq:temperature_field}
  \end{equation}
where, $\bm{\bar{v}}$ denotes the Darcy velocity vector and $T_0$ denotes the bulk fluid temperature, respectively. 

The governing equations of the two-layer DF model are then derived, consisting of the continuity equation, momentum equation and energy equations for the thermal-fluid and base-solid layers expressed as: 

%
%
  \begin{equation}
    \nabla \cdot \bm{\bar{v}}=0
    \label{eq:Continuity_equation}
  \end{equation}
  \begin{equation}
    \frac{6}{5} \rho_\mathrm{f} (\bm{\bar{v}} \cdot \nabla)\bm{\bar{v}}=-\nabla p + \mu \nabla ^2 \bm{\bar{v}} - \alpha(\gamma_1,\gamma_2)\bm{\bar{v}} -\beta(\gamma_1,\gamma_2)|\bm{\bar{v}}|\bm{\bar{v}}
    \label{eq:Momentum_equation}
  \end{equation}
  \begin{equation}
    \rho_\mathrm{f} c_\mathrm{pf} \bm{\bar{v}} \cdot \nabla T_0=k(\gamma_1,\gamma_2) \nabla^2 T_0 + \frac{h(\gamma_1,\gamma_2)}{2H_\mathrm{t}}(T_\mathrm{b0}-T_0)
    \label{eq:thermal-fluid_layer}
  \end{equation}
  \begin{equation}
    k_\mathrm{s} \nabla^2 T_\mathrm{b0}-\frac{h(\gamma_1,\gamma_2)}{2H_\mathrm{b}}(T_\mathrm{b0}-T_0)+\frac{q_\mathrm{s}}{2H_\mathrm{b}}=0
    \label{eq:base-solid_layer}
  \end{equation}
where $T_\mathrm{b0}$ denotes the base temperature at the central plane. The detail derivation of the equations is shown in the \ref{app1}. \par
The Darcy coefficient $\alpha$ and the Forchheimer coefficient $\beta$, representing the viscous and inertial resistances in porous media, are defined according to the DF law (Eq.~\eqref{eq:DFlaw}), and the effective interfacial heat transfer coefficient is expressed by Eqs. \eqref{eq:htc_thermalfluid}--\eqref{eq:htc_total} as: 
  \begin{equation}
    \nabla p=-\frac{\mu}{\kappa}\bm{\bar{v}}-\frac{\rho_\mathrm{f} c_\mathrm{F}}{\sqrt{\kappa}}|\bm{\bar{v}}|\bm{\bar{v}}=- \alpha(\gamma_1,\gamma_2)\bm{\bar{v}} -\beta(\gamma_1,\gamma_2)|\bm{\bar{v}}|\bm{\bar{v}}
    \label{eq:DFlaw}
  \end{equation}

  \begin{equation}
    h_\mathrm{t}=\frac{35 k(\gamma_1,\gamma_2)}{26H_\mathrm{t}}
    \label{eq:htc_thermalfluid}
  \end{equation}
  \begin{equation}
    h_\mathrm{b}=\frac{k_\mathrm{s}}{H_\mathrm{b}}
    \label{eq:htc_basesolid}
  \end{equation}

  \begin{equation}
    h(\gamma_1,\gamma_2)=\frac{h_\mathrm{t} h_\mathrm{b}}{h_\mathrm{t}+h_\mathrm{b}}
    \label{eq:htc_total}
  \end{equation}


%


\subsection{Interpolation of equivalent physical properties}
\label{subsec2.3}

The physical properties of lattice are subsequently derived as a function of the design variable $\gamma_2$. 
Relative density $\rho_\mathrm{por}$ is defined as the ratio of the volume of the solid material to the total unit cell volume and expressed as: 
\begin{equation}
  \centering
  \rho_\mathrm{por}=\frac{V_\mathrm{s}}{V_\mathrm{tot}}=1-\epsilon_\mathrm{por}
\end{equation}
where, $\epsilon_\mathrm{por}$ denotes the porosity of lattice.\par
\begin{figure}[htbp]
    \centering
    \begin{subfigure}{0.45\linewidth}
          \centering
          \includegraphics[height=0.7\linewidth]{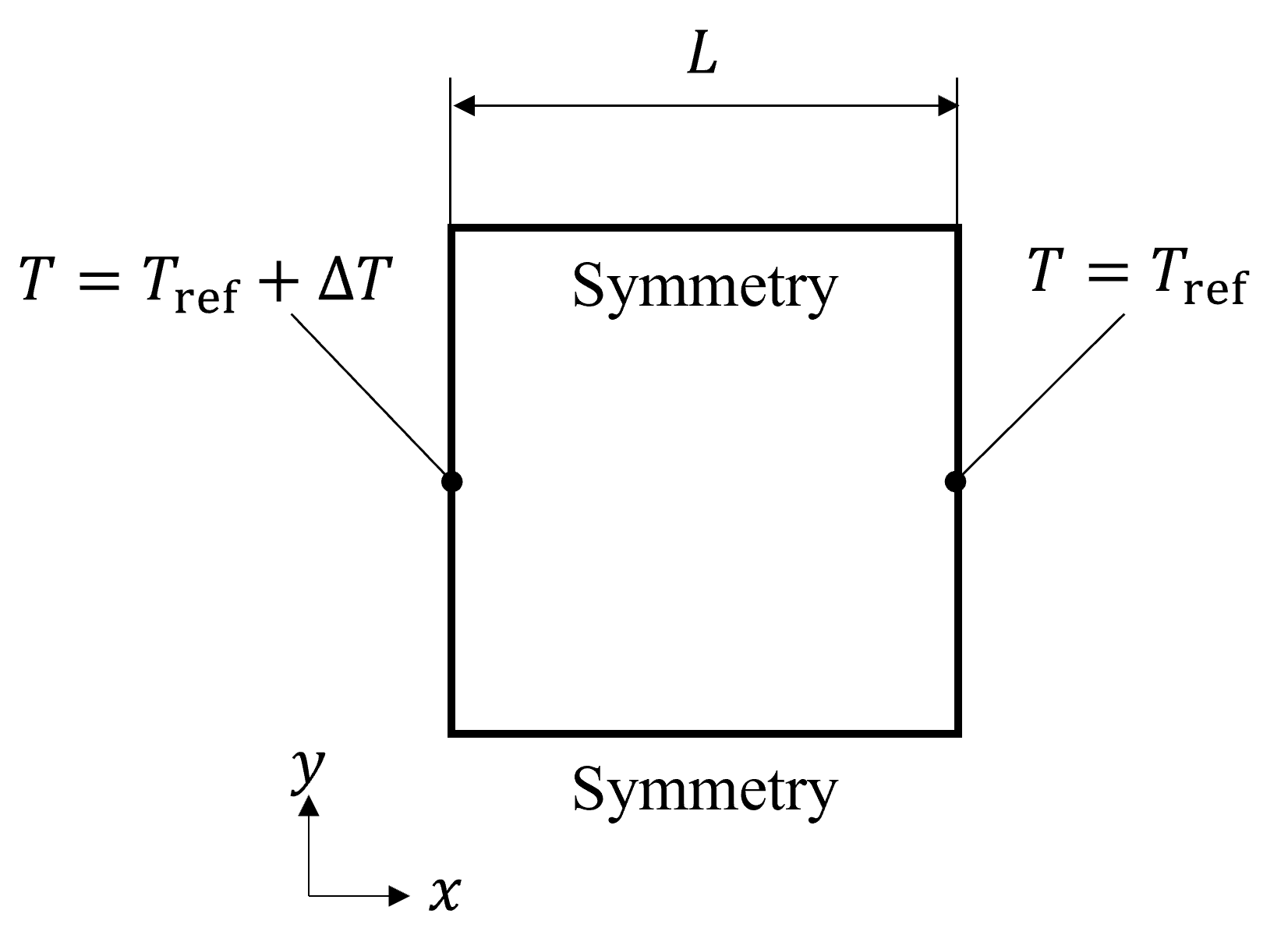}
          \caption{}
          \label{fig:RVE_ht}
      \end{subfigure}
      \hfill
      \begin{subfigure}{0.45\linewidth}
          \centering
          \includegraphics[height=0.7\linewidth]{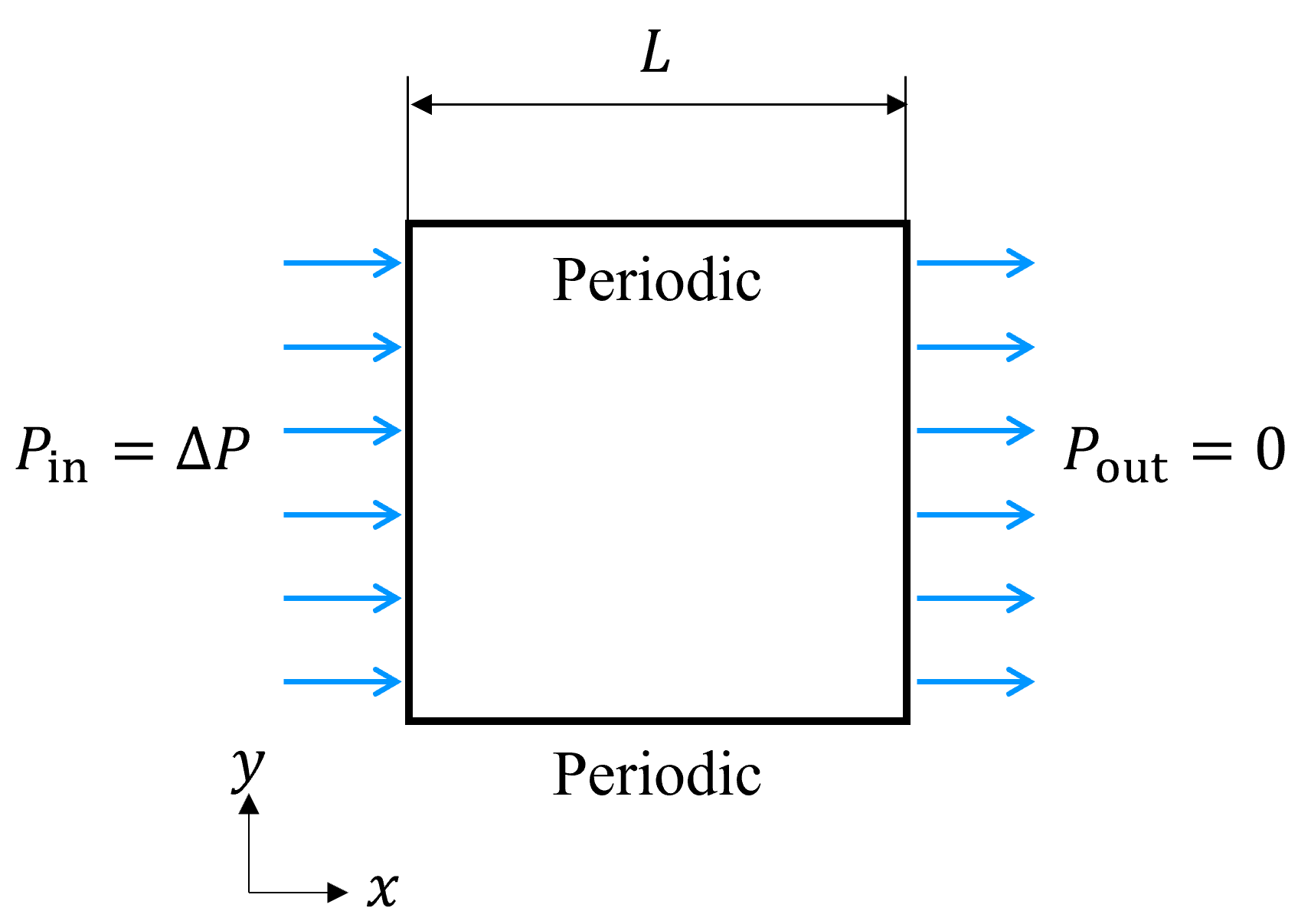}
          \caption{}
          \label{fig:RVE_pd}
      \end{subfigure} 
      \caption{Boundary condition of RVE domain for (a) heat conductivity, (b) fluid resistance. }
      \label{fig:RVE_method}
  \end{figure}

The equivalent thermal conductivity and flow resistance coefficients are derived using the representative volume element (RVE) method [32,33]. 
The boundary conditions for RVE domains is shown in \figref{RVE_method}. \par
Effective thermal conductivity, denoted as $k_\mathrm{por}$, is obtained by 
via Fourier\textquoteright s law:
  \begin{equation}
    k_\mathrm{por}=\frac{qL}{\Delta T}
    \label{eq:fourier's law}
  \end{equation}
and the dimensionless conductivity, denoted as $m_k(\gamma_2)$, is defined as: 
  \begin{equation}
    m_\mathrm{k}(\gamma_2)=\frac{k_\mathrm{por}-k_\mathrm{f}}{k_\mathrm{s}-k_\mathrm{f}}
    \label{eq:dimensionlesshcc}
  \end{equation}

\begin{figure}[htbp]
    \centering
    \includegraphics[width=0.8\linewidth]{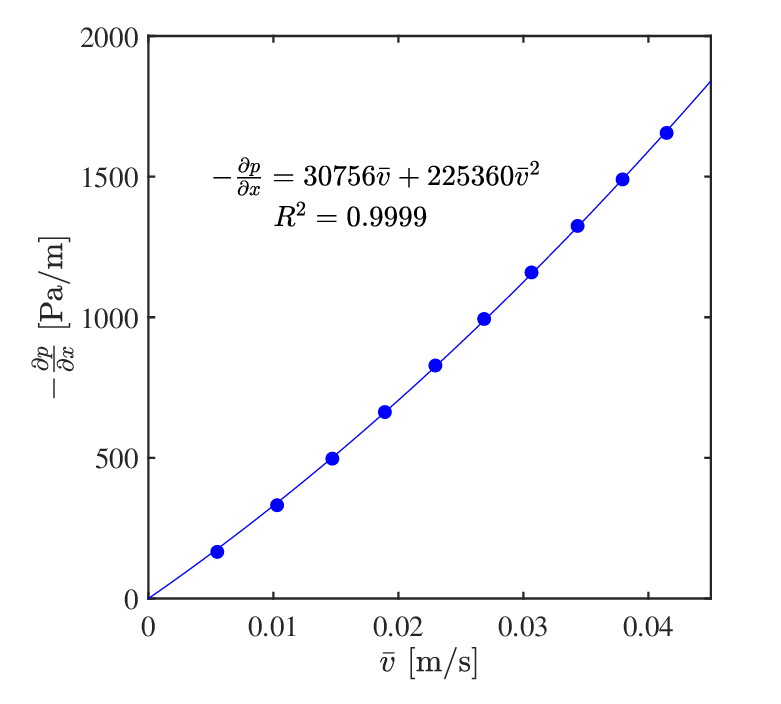}
    \caption{Interpolation function of pressure gradient with respect to Darcy velocity of the lattice. }
    \label{fig:dv-pd_of_0.6}
  \end{figure}

While calculating fluid resistance coefficients, several inlet pressure conditions are imposed. 
Then, relationship between darcy velocity and pressure gradient $-\partial p/\partial x$ (=$\Delta P/L$) is obtained. 
The pressure drop gradient $-\partial p/\partial x$ is approximated by a quadric order polynomial with respect to the darcy velocity $\bar{v}$ based on DF theory (Eq.~\eqref{eq:DFlaw}). 
\figref{dv-pd_of_0.6} represents an example of relationship between magnitude of darcy velocity $\bar{v}$ and $-\partial p/\partial x$. 
In this case, Darcy coefficient $\alpha_\mathrm{por}$ and Forchheimer coefficient $\beta_\mathrm{por}$ are derived as 30756 and 225360 respectively.

Finally, the obtained equivalent porosity is linearly interpolated, whereas other properties are interpolated using an extended RAMP scheme, as given by Eqs.~\eqref{eq:porosity}--\eqref{eq:beta}: 
  \begin{equation}
    \epsilon(\gamma_1,\gamma_2)=1+(\epsilon_\mathrm{por}(\gamma_2)-1)(1-\gamma_1)
    \label{eq:porosity}
  \end{equation}
  \begin{equation}
    k(\gamma_1, \gamma_2)=k_\mathrm{f}+(k_\mathrm{por}(\gamma_2)-k_\mathrm{f})\frac{1-\gamma_1}{1+q_\mathrm{k} \gamma_1}
    \label{eq:hcc}
  \end{equation}
  \begin{equation}
    \alpha(\gamma_1,\gamma_2)=\alpha_\mathrm{f}+(\alpha_\mathrm{por}(\gamma_2)-\alpha_\mathrm{f})\frac{1-\gamma_1}{1+q_\mathrm{f} \gamma_1}
    \label{eq:alpha}
  \end{equation}
  \begin{equation}
    \beta(\gamma_1,\gamma_2)=\beta_\mathrm{f}+(\beta_\mathrm{por}(\gamma_2)-\beta_\mathrm{f})\frac{1-\gamma_1}{1+q_\mathrm{f} \gamma_1}
    \label{eq:beta}
  \end{equation}
Here, $\alpha_\mathrm{f}$ and $\beta_\mathrm{f}$ denote the viscous and inertial resistance parameters of the fluid in the absence of porous media, with $\alpha_\mathrm{f}$ set to $3\mu_\mathrm{f}/H_\mathrm{t}^2$ and $\beta_\mathrm{f}$ set to 0, respectively. 
Eqs. \eqref{eq:porosity}--\eqref{eq:beta} show that the region of $\gamma_1=1$ represent fluid properties, whereas $\gamma_1=0$ region represent those of porous.

\subsection{Optimization problem setting}
\label{subsec2.4}
The optimization problem is formulated with the objective of minimizing the maximum base-plate temperature. 
In this study, the $p$-norm approach is adopted to achieve this, expressed as: 
  \begin{equation*}
    \mathrm{Minimize} \: \left[\frac{1}{A_{\Omega_\mathrm{d}}} \int_{\Omega_\mathrm{d}}(T_\mathrm{b0}-T_\mathrm{in})^p d\Omega \right]^{\frac{1}{p}} 
  \end{equation*}
  \begin{equation}
    \mathrm{Subject\:to} \: \int_{\Omega_\mathrm{d}}(1-\epsilon(\gamma_1,\gamma_2)) d\Omega \leq  f A_{\Omega_\mathrm{d}}
    \label{eq:Objective}
  \end{equation}
    \begin{equation*}
    0 \leq \gamma_1,\gamma_2 \leq 1
  \end{equation*} 
where, $\Omega_\mathrm{d}$ denotes design domain and $A_{\Omega_\mathrm{d}}$ indicates its area. 
Following previous studies \cite{Yanetal,Fromyaji}, a value of $p=10$ is employed. 
Unless otherwise specified, a trivial volume constraint is applied by setting 
$f=1$, which is equivalent to the case without an explicit volume constraint.

\subsection{Numerical implementation}
\label{subsec2.5}
\begin{figure}[htbp]
      \centering
      \includegraphics[width=\linewidth]{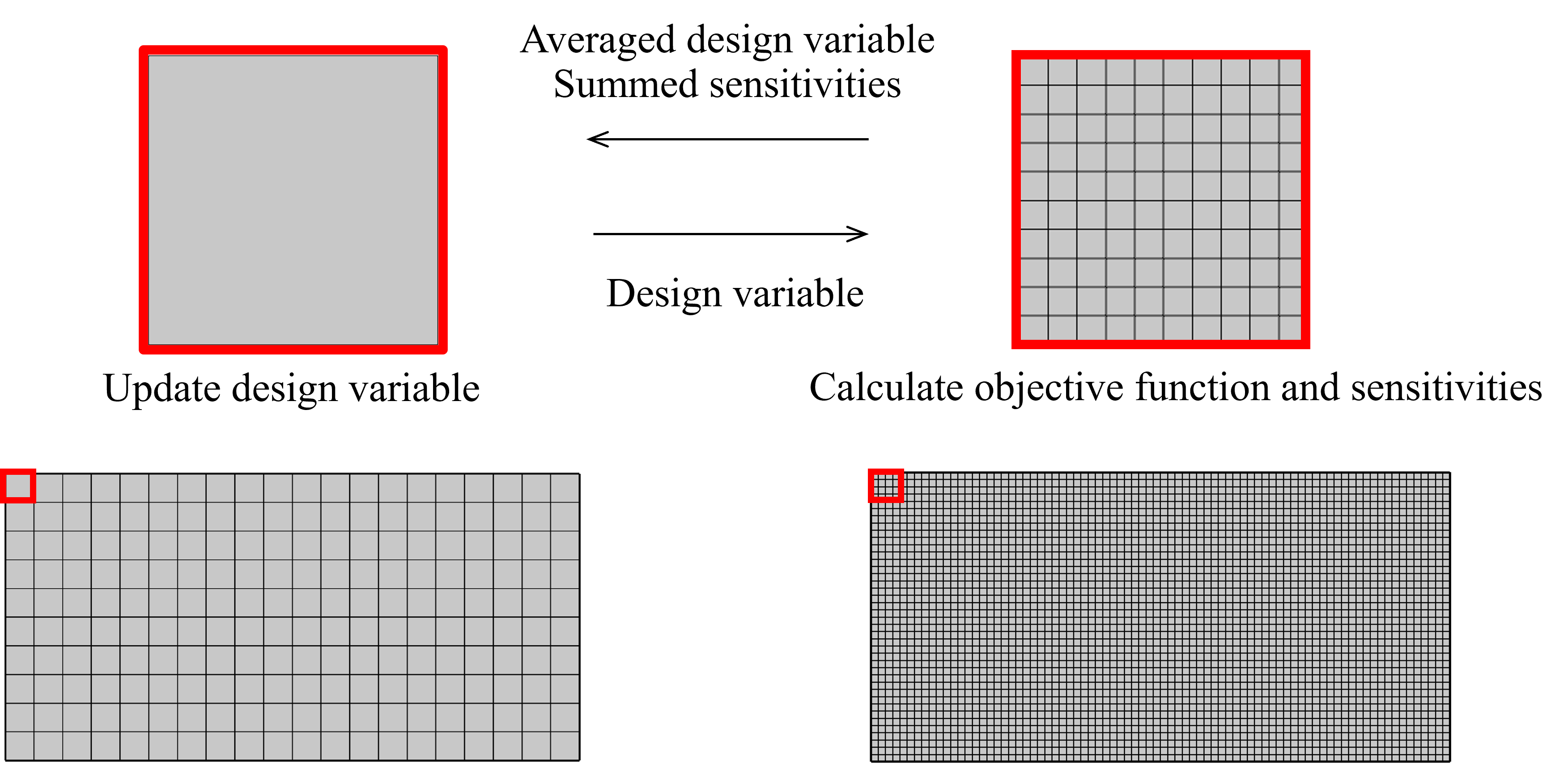}
      \caption{Schematic of density mapping algorithm through updating process. }
      \label{fig:Density_mapping}
  \end{figure}

In this study, finite element (FE) calculations are performed using commercial FE software, COMSOL Multiphysics 6.2 with laminar flow and heat transfer modules. 
Sensitivities are calculated by adjoint method, using sensitivity module in COMSOL. 
The optimization algorithm is implemented using the method of moving asymptotes (MMA) \cite{MMA_theory}, 
with the design variables updated at each iteration in MATLAB.
In the traditional topology optimization, design variables are defined in each FE mesh.
On the other hand, a density mapping scheme is applied in present study to define design variables for each unit. 
To ensure consistent updating of design variables and sensitivities within each unit, 
the design variables and sensitivities within each unit are averaged and summed, respectively in each iteration step. After updating design variables, the values are redistributed across each discretized unit cell (see \figref{Density_mapping}).
This process also enables us to perform FE analysis with sufficient fine mesh setting regardless of unit size.

The initial values of design variables $\gamma_1$ and $\gamma_2$ are both set to 0 and the optimization is conducted for 200 iterations. 
To promote binarization of porous and void phases, a continuation approach \cite{Yanetal,NaturalconvTO,NabeLKS} is applied in which the convexity parameters are varied across iterations.
Specifically, convexity parameter for the effective conductivity, $q_\mathrm{k}$ (see Eq.~\eqref{eq:hcc}) is gradually increased (1, 5, 10, 50), 
while convexity parameter for the effective fluid resistance, $q_\mathrm{f}$ (see Eq.~\eqref{eq:alpha} and~\eqref{eq:beta})
is gradually decreased (50, 10, 5, 1), with updates performed every 50 design iterations. \par
Heaviside projection \cite{projection,Mnd} is also applied as:  
  \begin{equation}
    \hat{\gamma_1}=\frac{\tanh(\beta \eta) + \tanh(\beta (\gamma_1-\eta))}{\tanh(\beta \eta) + \tanh(\beta (1-\eta))}
    \label{eq:projection}
  \end{equation} 
Unless otherwise stated, steepness parameter $\beta$ is set to 1, while threshold $\eta$ is set to 0.5. 
The parameter values described in this subsection are determined empirically to achieve a reasonable balance between accuracy and numerical stability.

\section{Results and discussion}
\label{sec3}

\subsection{Problem setting}
\label{subsec3.1}
\subsubsection{The property of lattice unit cell}
\label{subsubsec3.1.1}
\begin{figure}[htbp]
    \centering
    \includegraphics[width=0.5\linewidth]{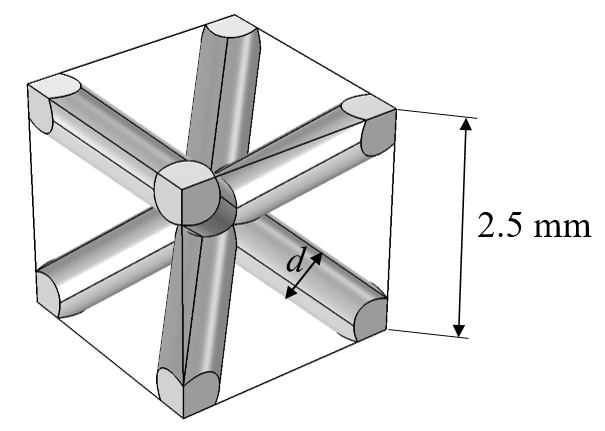}
    \caption{BCC lattice unit cell. }
    \label{fig:BCClattice}
  \end{figure}

\begin{figure}[htbp]
      \begin{subfigure}{0.5\linewidth}
          \centering
          \includegraphics[width=\linewidth]{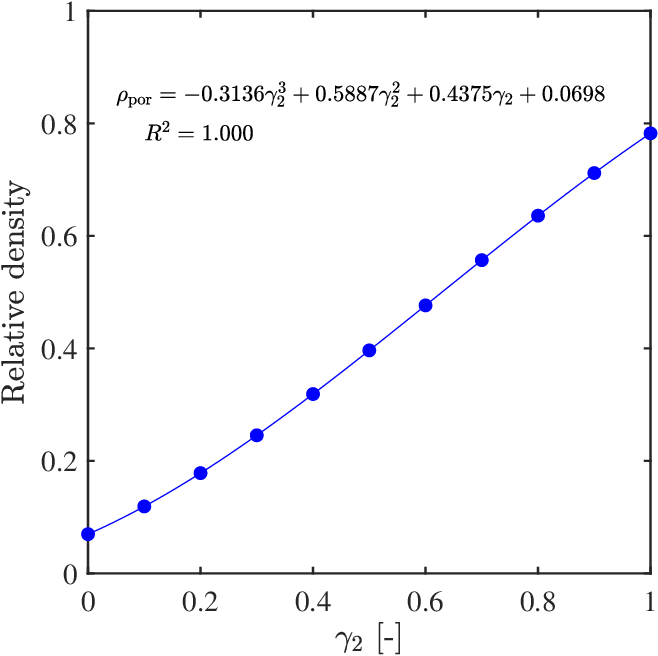}
          \caption{}
          \label{fig:RVE_reden}
      \end{subfigure}
      \hspace{0.05\linewidth}
      \begin{subfigure}{0.5\linewidth}
          \centering
          \includegraphics[width=\linewidth]{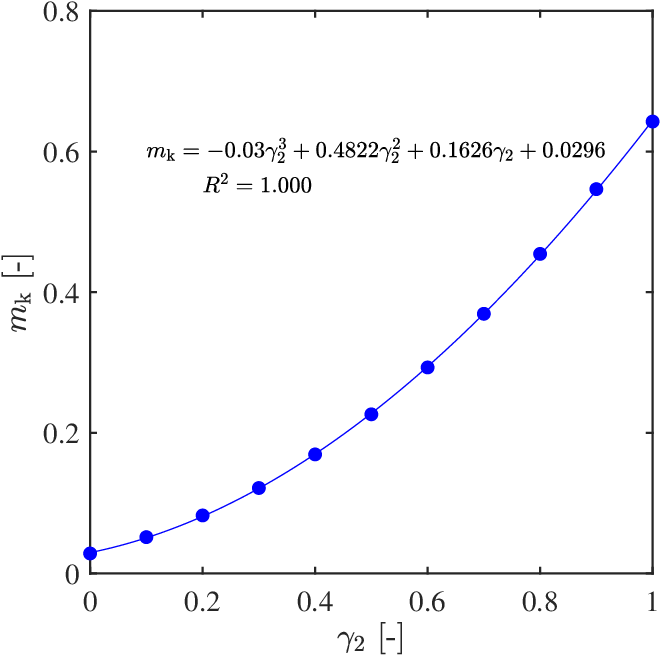}
          \caption{}
          \label{fig:RVE_ht_prop}
      \end{subfigure}
      
      \vspace{1em} 
      
      \begin{subfigure}{0.5\linewidth}
          \centering
          \includegraphics[width=\linewidth]{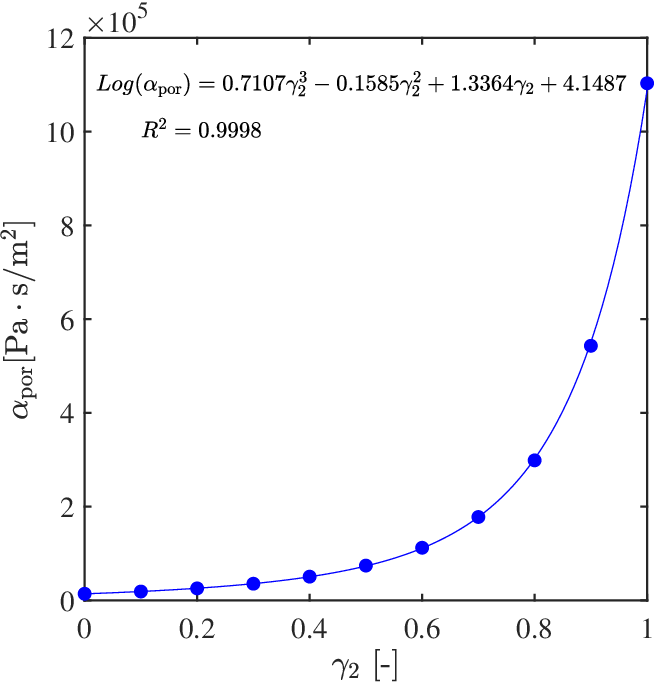}
          \caption{}
          \label{fig:RVE_alpha}
      \end{subfigure}
      \hspace{0.05\linewidth}
      \begin{subfigure}{0.5\linewidth}
          \centering
          \includegraphics[width=\linewidth]{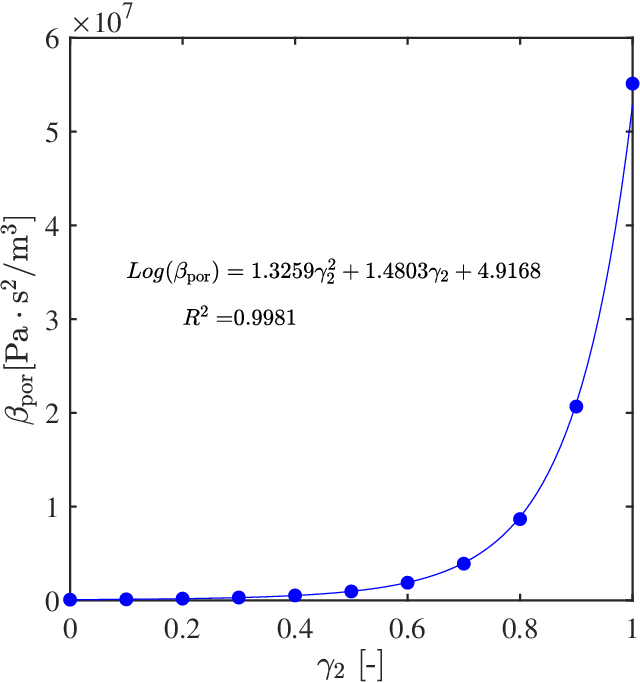}
          \caption{}
          \label{fig:RVE_beta}
      \end{subfigure}
      
      \caption{Interpolation function of (a) Relative density, (b) Dimensionless heat conductivity, (c) Darcy coefficient, (d) Forchheimer coefficient. }
      \label{fig:RVE_property}
  \end{figure}

In this study, the lattice unit is modeled as a body-centered cubic (BCC) structure (\figref{BCClattice}), which is self-supported and feasible to manufacture via powder bed fusion (PBF) manufacturing process \cite{Ushijimalattice,Kosakalatice,Takarazawa}. 
The unit size is set to 2.5 mm. 
The lattice beam diameter is constrained within 0.3--1.3 mm to ensure its manufacturability ($d_\mathrm{min}=0.3\: \mathrm{mm}$, $d_\mathrm{max}=1.3\: \mathrm{mm}$). 
The equivalent properties of lattice unit are interpolated as \figref{RVE_property}. 

\subsubsection{The analysis setting}
\label{subsubsec3.1.2}
\begin{figure}[htbp]
    \centering
      \begin{subfigure}{0.8\linewidth}
          \centering
          \includegraphics[width=\linewidth]{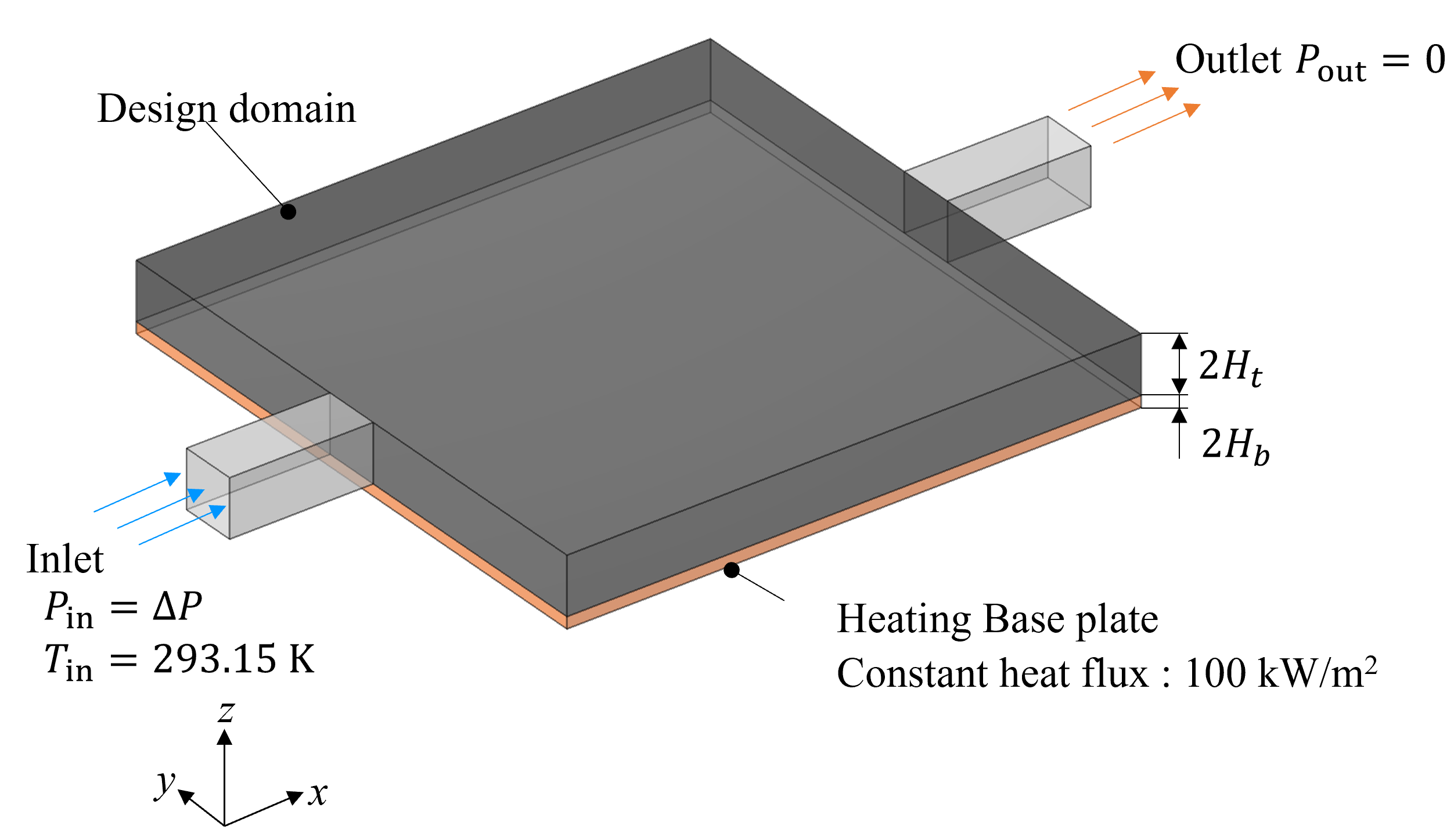}
          \caption{}
          \label{fig:3Dfull}
      \end{subfigure}
      \begin{subfigure}{0.6\linewidth}
          \centering
          \includegraphics[width=\linewidth]{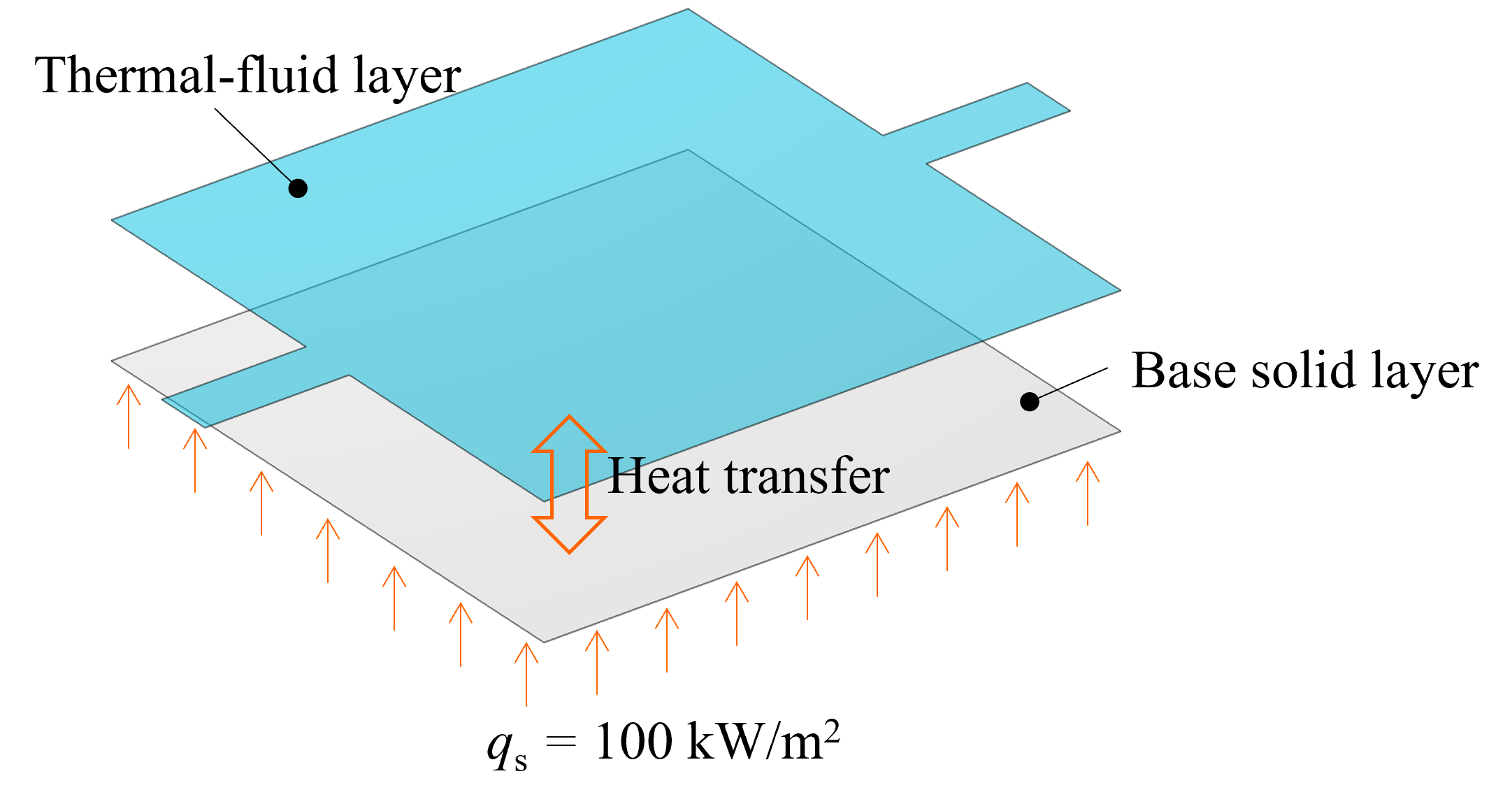}
          \caption{}
          \label{fig:2layer}
      \end{subfigure}
      \begin{subfigure}{0.6\linewidth}
          \centering
          \includegraphics[width=\linewidth]{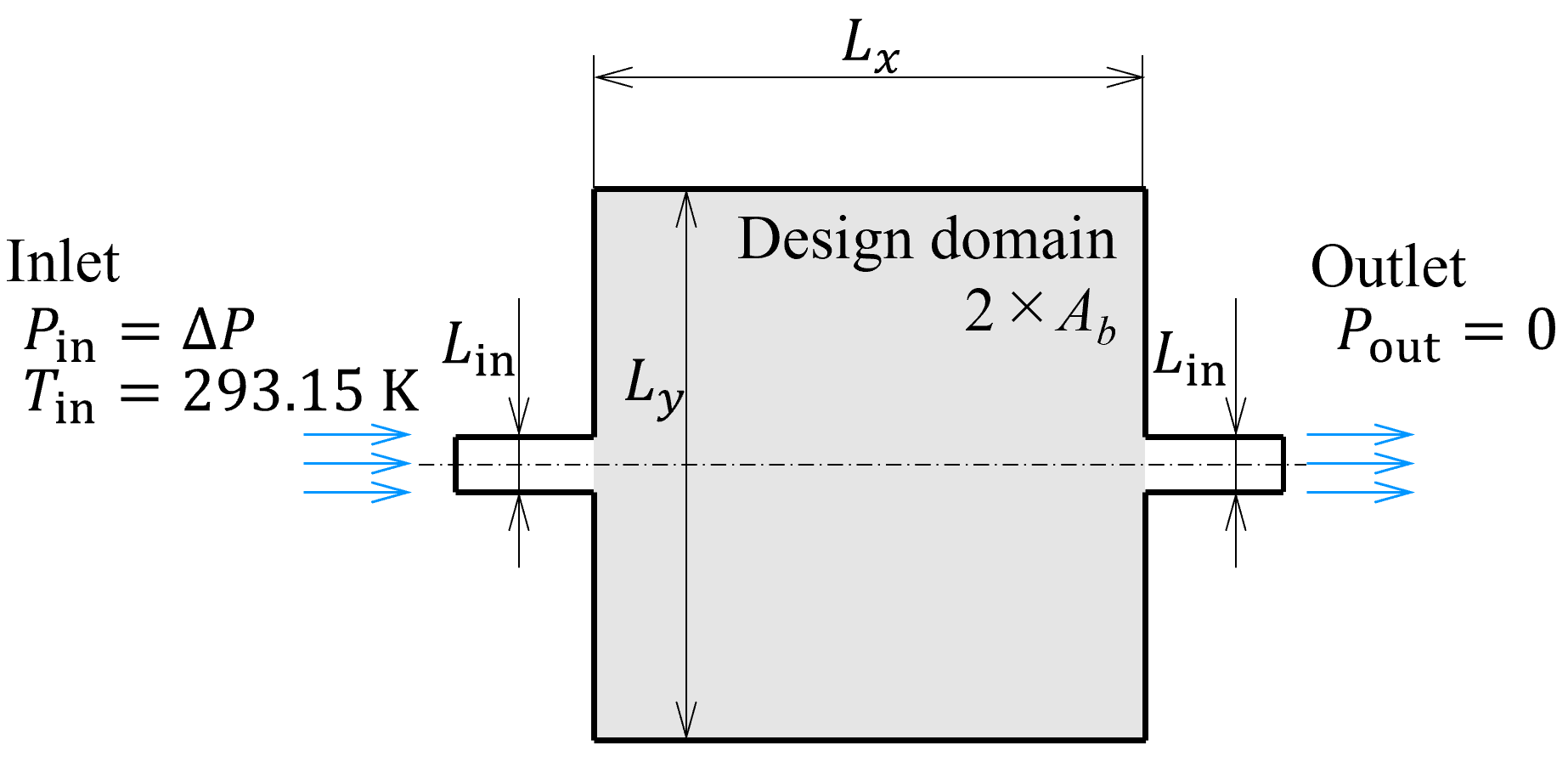}
          \caption{}
          \label{fig:2Dmodel}
      \end{subfigure}  
      \caption{The water cooled heat sink domain for optimization (a) 3D-full model. (b) Two-layer design domain. (c) Boundary condition and dimensions of 2D model. }
      \label{fig:Boundary_condition}
  \end{figure}

The 3D computational domain and boundary conditions are illustrated in \figref{3Dfull}. 
The total size of design domain is $50 \times 50 \times 5$ mm. Thus, at most $20 \times 20 \times 2$ unit cells can be placed inside. 
The pressure is fixed to the constant value $P_\mathrm{in}=\Delta P$ and temperature is fixed to $T_\mathrm{in}=293.15  \; \mathrm{K}$ at inlet, 
while the pressure is fixed to $P_\mathrm{out}=0 \: \mathrm{Pa}$ at outlet. 
A uniformly distributed heat flux $q_\mathrm{s}=100 \: \mathrm{kW/m^2}$ is applied to the bottom surface of the base plate.
The problem is reformulated using the two-layer DF model, as depicted in 
\figref{2layer} and \figref{2Dmodel} to reduce computational cost. 
Furthermore, symmetry conditions allow the use of a half-domain model. 
The dimensions of the Heat sink and physical properties of both the fluid and solid phases are summarized in \tabref{model_setting}. 
%
%
%
%
  \begin{table}[htbp]
    \centering
    \caption{Dimensions of the model and material property. }
    \begin{tabular}{cc}
      \hline
      \multicolumn{2}{c}{Dimensions of the model} \\ \hline

      $L_x \; \mathrm{[mm]}$                                  & 50 \\ 
      $L_y  \; \mathrm{[mm]}$                                  & 50 \\
      Inlet size $L_\mathrm{in}  \; \mathrm{[mm]} \times L_\mathrm{in}  \; \mathrm{[mm]}$                  & 5  \\ 
      Thickness of thermal-fluid layer 2$H_\mathrm{t} \; \mathrm{[mm]}$ & 5  \\ 
      Thickness of base solid layer 2$H_\mathrm{b} \; \mathrm{[mm]}$    & 1  \\ \hline

      \multicolumn{2}{c}{Material property of fluid \: (water)} \\ \hline
      Viscosity coefficient $ \mu_\mathrm{f}  \; \mathrm{[Pa \cdot s]}$	& $1.004\times10^{-3}$ \\ 
      Density $\rho_\mathrm{f}  \; \mathrm{[kg/m^3]}$	& 998 \\ 
      Heat conductivity $k_\mathrm{f} \; \mathrm{[W/(m \cdot K)]}$ &	0.598 \\ 
      Specific heat $c_\mathrm{pf} \; \mathrm{[J/(kg \cdot K)]}$ &	4180 \\ \hline
      
      \multicolumn{2}{c}{Material property of solid \: (aluminum alloy)} \\ \hline
      Density $\rho_\mathrm{s} \; \mathrm{[kg/m^3]}$	& 2000 \\ 
      Heat conductivity $k_\mathrm{s} \; \mathrm{[W/(m \cdot K)]}$ &	100 \\ 
      Specific heat $c_\mathrm{ps} \; \mathrm{[J/(kg \cdot K)]}$ &	900 \\ \hline
    \end{tabular}
    \label{tab:model_setting}
  \end{table}

\subsection{Validation of the two-layer DF model using random distributions}
\label{subsec3.2}

In this study, following approach is applied to evaluate the validity of employing the two-layer DF model for optimization. 
Here, thirty sample designs are generated, each consisting of randomly distributed design variables ($\gamma_1$,$\gamma_2$). 
The variable $\gamma_1$ is binarized using a threshold of 0.5 to explicitly distinguish between void and lattice regions. 
The probability of void occurrence is randomly set within the range of 0.2--0.8 to ensure computational stability, 
thereby producing samples ranging from sparse to dense lattice distributions. 
  \begin{figure}[htbp]
      \begin{subfigure}{0.5\linewidth}
          \centering
          \includegraphics[width=\linewidth,draft=false]{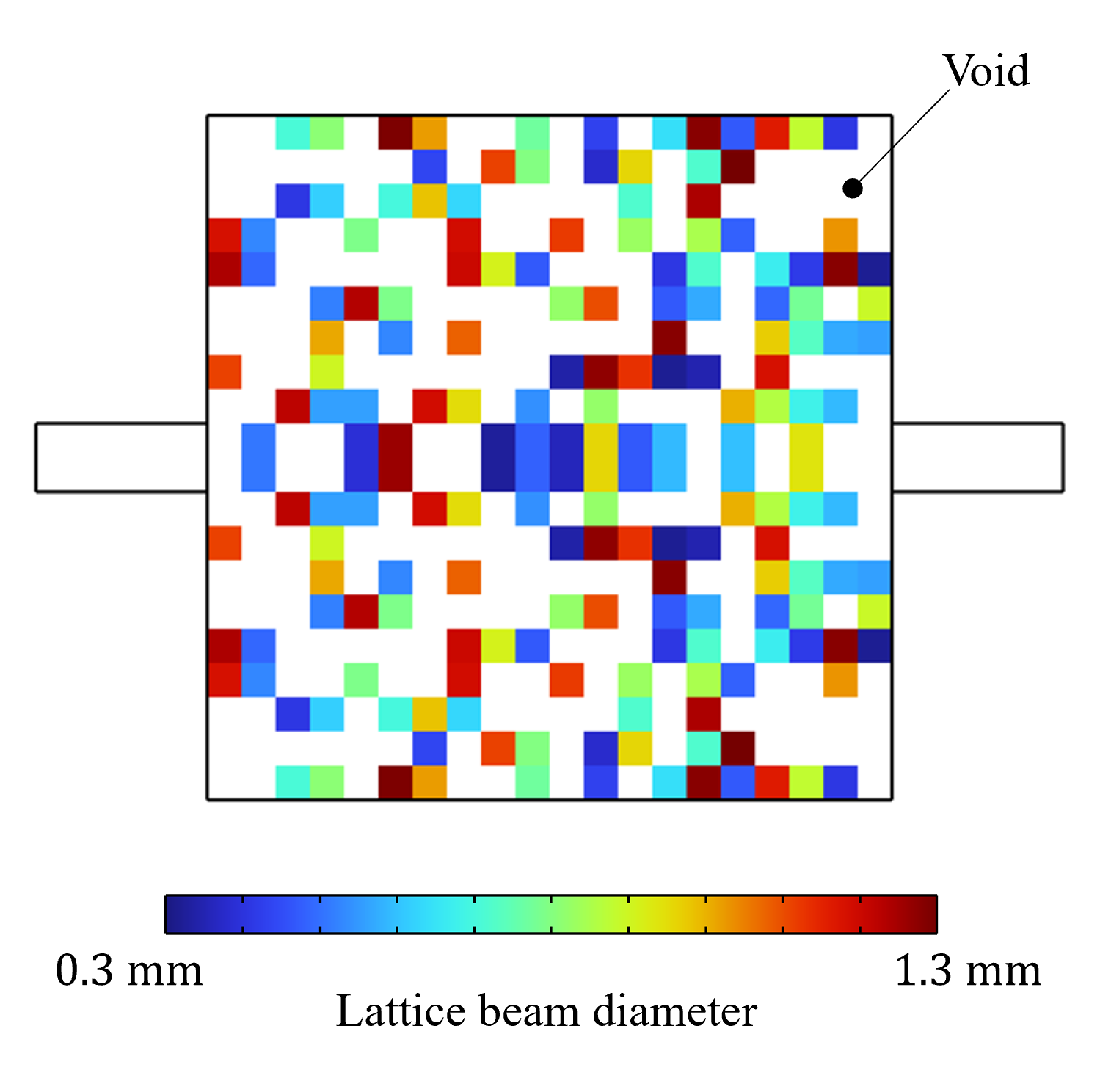}
          \caption{}
          \label{fig:porous_rand_example}
      \end{subfigure}
      \hspace{0.05\linewidth}
      \begin{subfigure}{0.45\linewidth}
          \centering
          \includegraphics[width=\linewidth]{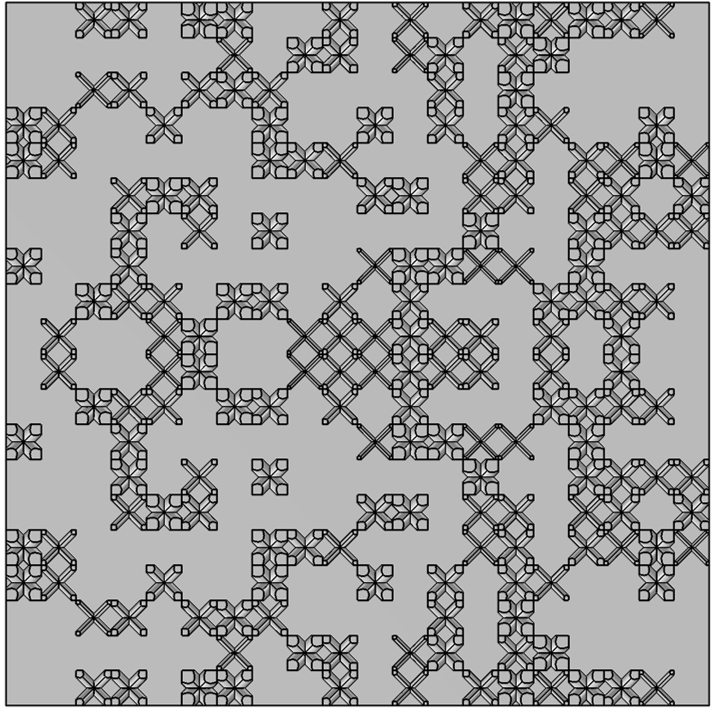}
          \caption{}
          \label{fig:FS_rand_example}
      \end{subfigure}
      
      \caption{Example of randomly distributed model: (a) Distribution of lattice beam diameter, (b) Coresponding full-scale geometry. }
      \label{fig:example_rand}
  \end{figure}
  \begin{figure}[htbp]
    \centering
    \includegraphics[width=\linewidth]{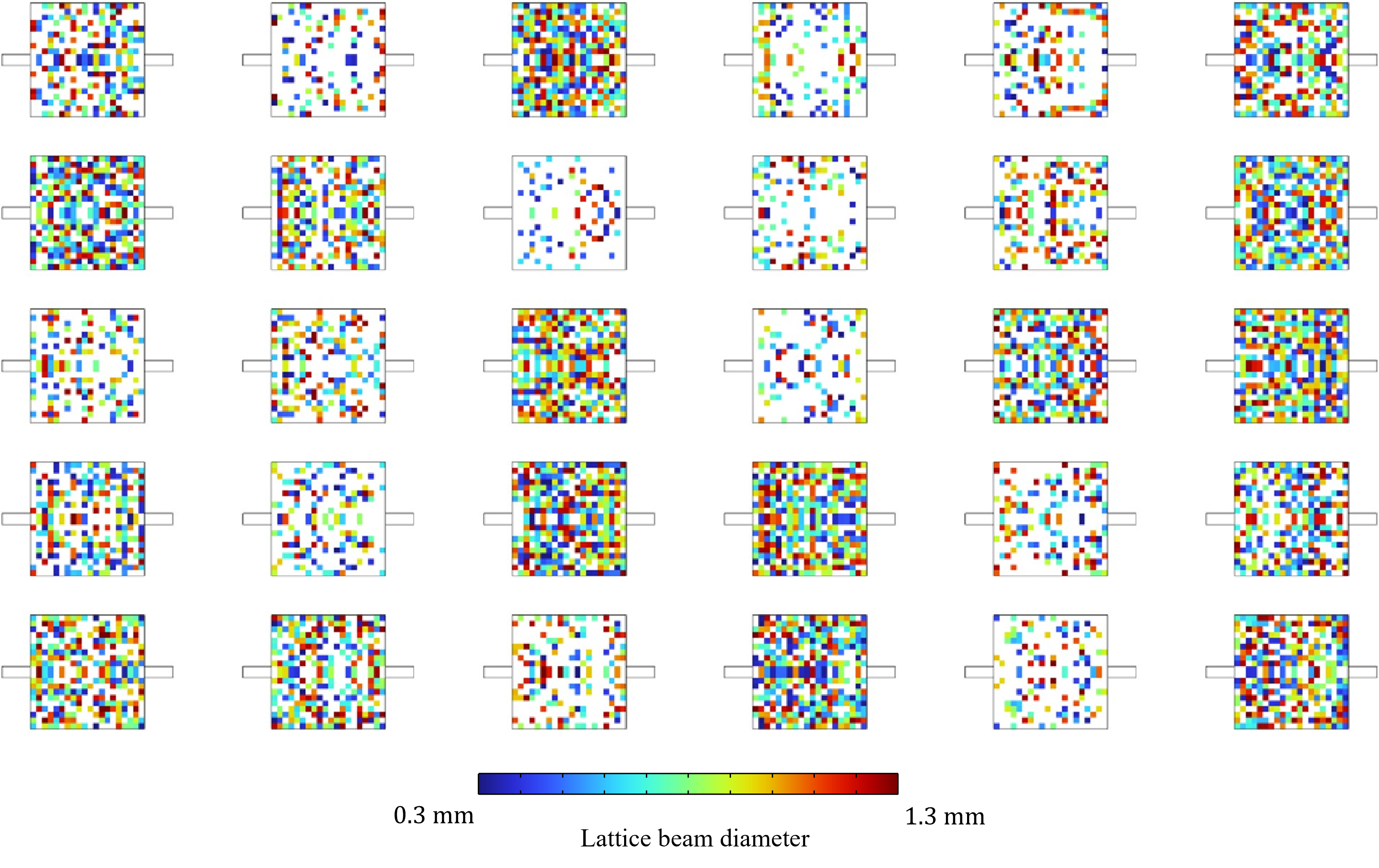}
    \caption{Illustration of the randomly distributed models used for validation. }
    \label{fig:randmodels}
  \end{figure}
Each sample is analyzed using both the reduced-order two-layer DF model and detailed full-scale analysis.
Furthermore, in this paper, several Nusselt number-based metrics are introduced to evaluate the thermal performance of the proposed structures.
These are defined in terms of local and averaged heat transfer coefficients, normalized by the fluid thermal conductivity $k_\mathrm{f}$ 
and the inlet characteristic length $L_\mathrm{in}$. \par
The maximum Nusselt number, $Nu_\mathrm{max}$ is defined using the maximum heat transfer coefficient $h_\mathrm{max}$ as:
  \begin{equation}
    Nu_{\mathrm{max}} = \frac{h_\mathrm{max} L_\mathrm{in}}{k_\mathrm{f}}
    \label{eq:Numax_def}
  \end{equation}
\par
Similarly, the objective Nusselt number, $Nu_\mathrm{obj}$ is defined using the objective heat transfer coefficient $h_\mathrm{obj}$ as:
  \begin{equation}
    Nu_{\mathrm{obj}} = \frac{h_\mathrm{obj} L_\mathrm{in}}{k_\mathrm{f}}
    \label{eq:Nuobj_def}
  \end{equation}
\par
In the same manner, the average Nusselt number, $Nu_\mathrm{ave}$, is defined using the average heat transfer coefficient $h_\mathrm{ave}$ as:
  \begin{equation}
    Nu_{\mathrm{ave}} = \frac{h_\mathrm{ave} L_\mathrm{in}}{k_\mathrm{f}}
    \label{eq:Nuave_def}
  \end{equation}
\par
The heat transfer coefficients in the above expressions are determined as follows. 
The maximum heat transfer coefficient, $h_\mathrm{max}$ corresponding to the maximum temperature at the base plate, is defined as:
  \begin{equation}
    h_\mathrm{max}=\frac{q}{T_\mathrm{bS,max}-T_\mathrm{in}}
    \label{eq:hmax_def}
  \end{equation}
where $q$ is the imposed heat flux, $T_\mathrm{bS,max}$ is the maximum temperature at the base-plate bottom surface, and $T_\mathrm{in}$ 
is the inlet fluid temperature.
\par
The objective heat transfer coefficient, associated with the optimization objective, is defined using the $p$-norm temperature distribution:
\begin{equation}
h_\mathrm{obj}=\frac{q}{[\frac{1}{A_\mathrm{b}} \int_{A_\mathrm{b}}(T_\mathrm{b0}-T_\mathrm{in})^p dA]^{\frac{1}{p}}}
\label{eq:hobj_def}
\end{equation}
where $T_\mathrm{b0}$ is the base temperature at the central plane and $A_\mathrm{b}$ is the base-plate area of the half-domain.
\par
Finally, the average heat transfer coefficient is expressed as:
\begin{equation}
h_\mathrm{ave}=\frac{q}{T_\mathrm{bS,ave}-T_\mathrm{in}}
\label{eq:have_def}
\end{equation}
where $T_\mathrm{bS,ave}$ is the average temperature at the bottom surface of the base plate. \par
\begin{figure}[htbp]
    \centering
    \begin{subfigure}{0.5\linewidth}
      \centering
      \includegraphics[height=\linewidth]{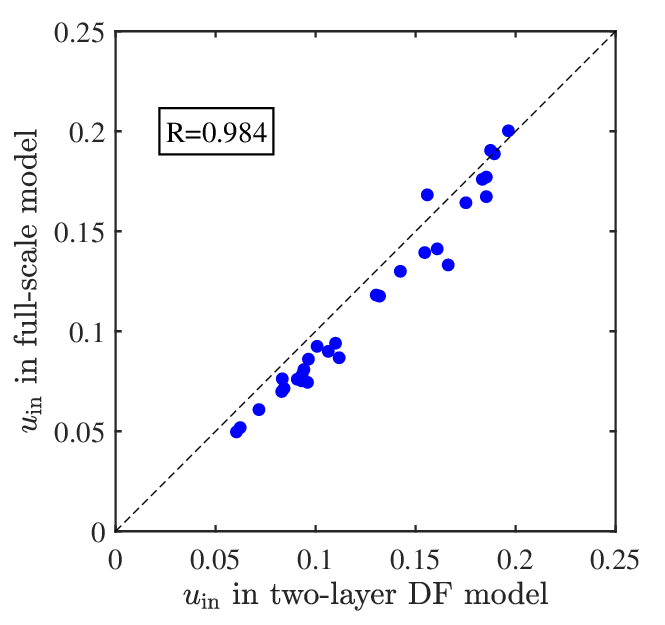}
      \caption{}
      \label{fig:Re_uin}
    \end{subfigure}\hfill
    \begin{subfigure}{0.5\linewidth}
      \centering
      \includegraphics[height=\linewidth]{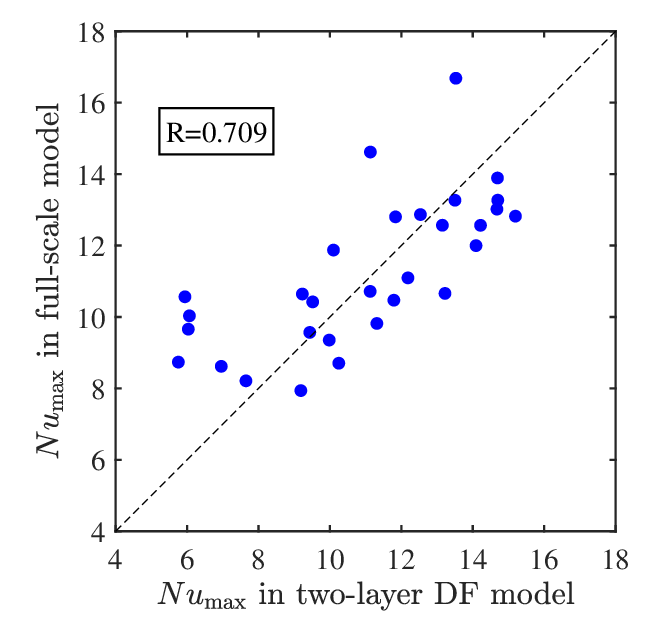}
      \caption{}
      \label{fig:Re_Numax}
    \end{subfigure}

    \vspace{0.75em}

    \begin{subfigure}{0.5\linewidth}
      \centering
      \includegraphics[height=\linewidth]{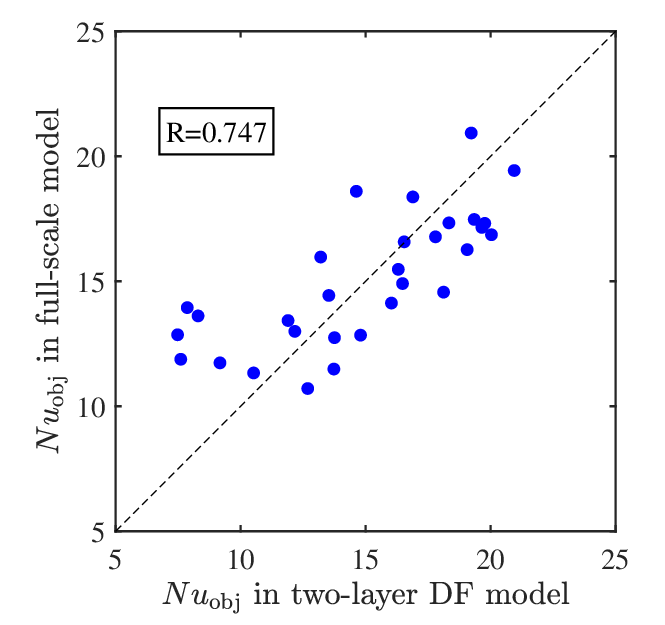}
      \caption{}
      \label{fig:Re_Nuobj}
    \end{subfigure}\hfill
    \begin{subfigure}{0.5\linewidth}
      \centering
      \includegraphics[height=\linewidth]{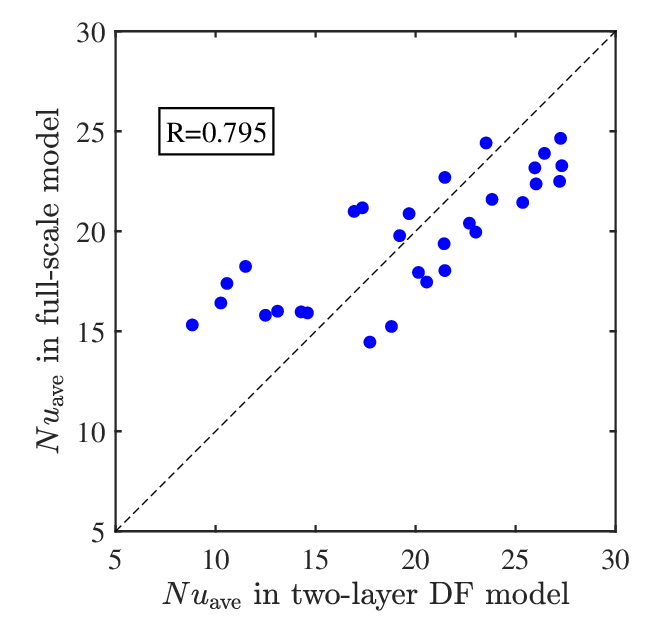}
      \caption{}
      \label{fig:Re_Nuave}
    \end{subfigure}
    \caption{Correlation between two-layer DF model and full-scale model: (a) $u_\mathrm{in}$, (b) $Nu_{\mathrm{max}}$, (c) $Nu_{\mathrm{obj}}$, (d) $Nu_{\mathrm{ave}}$. }
    \label{fig:Relations}
  \end{figure}

Since the performance of heat sinks is more appropriately evaluated based on the temperature at the heated surface, 
$h_\mathrm{max}$ and $h_\mathrm{ave}$ are calculated using temperature at the bottom surface of base plate. 
However, in the two-layer DF model, the temperature is defined at the central plane of the base plate. 
Since the thickness of base plate is thin enough, the temperature difference between the central plane and the bottom surface is estimated to be 0.5 K according to Fourier\textquoteright s law. 
This discrepancy is negligible compared with the overall temperature rise and thus has little influence on the calculated Nusselt numbers. 
\figref{Relations} compares the results obtained from the two-layer DF model and full-scale model. 
The correlation coefficient for inlet velocity exceeds 0.9, indicating strong agreement between the two models. 
Although the correlation for Nusselt numbers are lower, they remain above 0.7, which is sufficient to confirm a consistent trend. 
These results suggest that while the two-layer DF model does not fully replicate the high-fidelity model, 
it provides adequate accuracy to serve as a low-fidelity surrogate for optimization.
\subsection{Optimization result}
\label{subsec3.3}
\begin{figure}[htbp]
  \centering
  \begin{subfigure}[b]{0.3\textwidth}
    \centering
    \includegraphics[height=\linewidth]{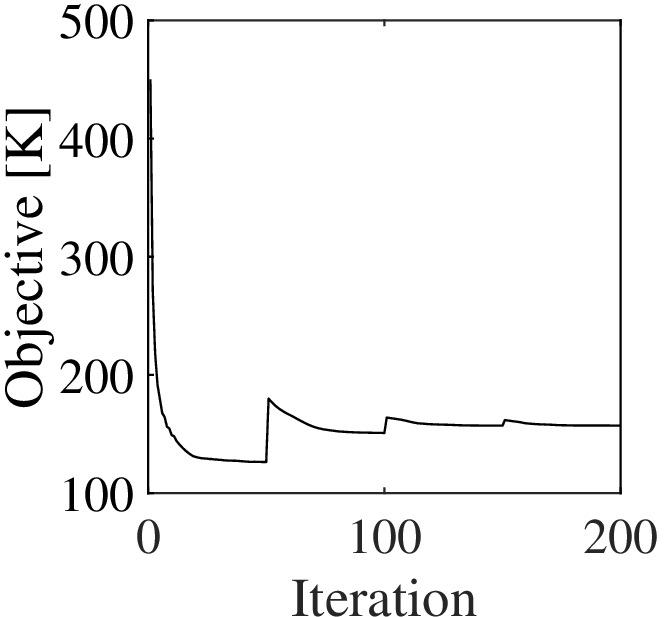}
    \caption{}
  \end{subfigure}
  \hfill
  \begin{subfigure}[b]{0.3\textwidth}
    \centering
    \includegraphics[height=\linewidth]{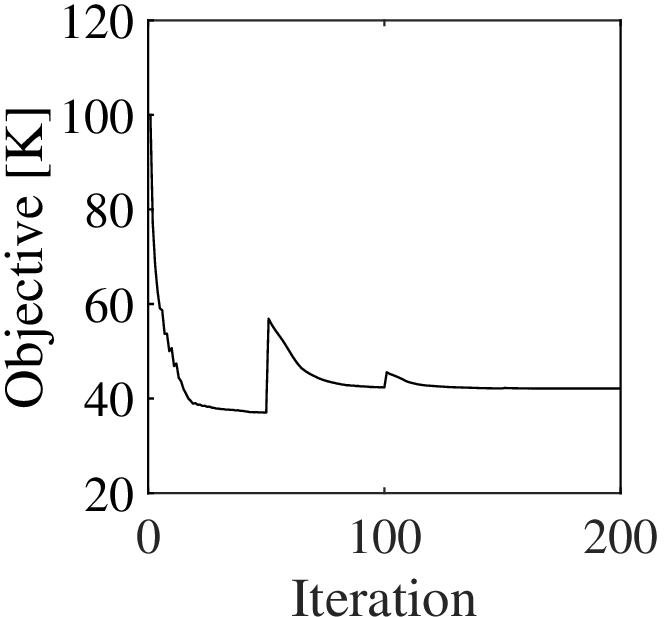}
    \caption{}
  \end{subfigure}
  \hfill
  \begin{subfigure}[b]{0.3\textwidth}
    \centering
    \includegraphics[height=\linewidth]{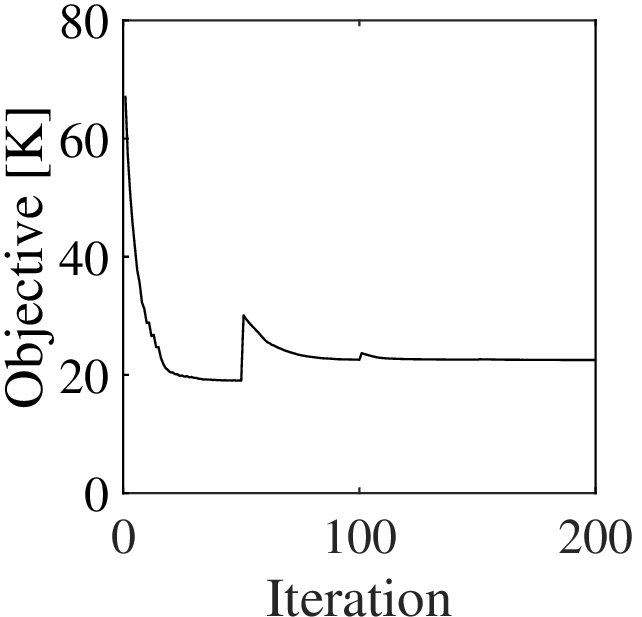}
    \caption{}
  \end{subfigure}%
  \caption{Optimization history of objective function: (a) $P_\mathrm{in}=$1 Pa, (b) $P_\mathrm{in}=$10 Pa, (c) $P_\mathrm{in}=$50 Pa. }
  \label{fig:Opt_hist}
\end{figure}

\figref{Opt_hist} illustrates the optimization histories of the objective function under three inlet pressure conditions 
($P_\mathrm{in}$ = 1 Pa, 10 Pa and 50 Pa).
Peaks observed in the plots is considered to result from the continuation approach.
In order to evaluate the degree of binarization of the design variable $\gamma_1$, we introduced a parameter $M_\mathrm{nd}$, defined by Eq.~\eqref{Mnd} \cite{Mnd} as follows:  
\begin{equation}
  \centering
  M_\mathrm{nd}=\sum_{e=1}^{n} \frac{4\gamma_{1,e}(1-\gamma_{1,e})}{n}
  \label{Mnd}
\end{equation}
\par
A smaller value of $M_\mathrm{nd}$ indicates that $\gamma_1$ approaches a 
binary state (0 or 1). 
The obtained values for $P_{\mathrm{in}}$=1 Pa, 
10 Pa, 50 Pa are 
0.931 \%, $4.99\times10^{-3}$ \% and 0.0389 \% respectively, 
confirming sufficient binarization in all cases.

\begin{figure}[htbp]
    \centering
    \begin{subfigure}[b]{0.3\textwidth}
        \centering
        \includegraphics[height=0.7\linewidth]{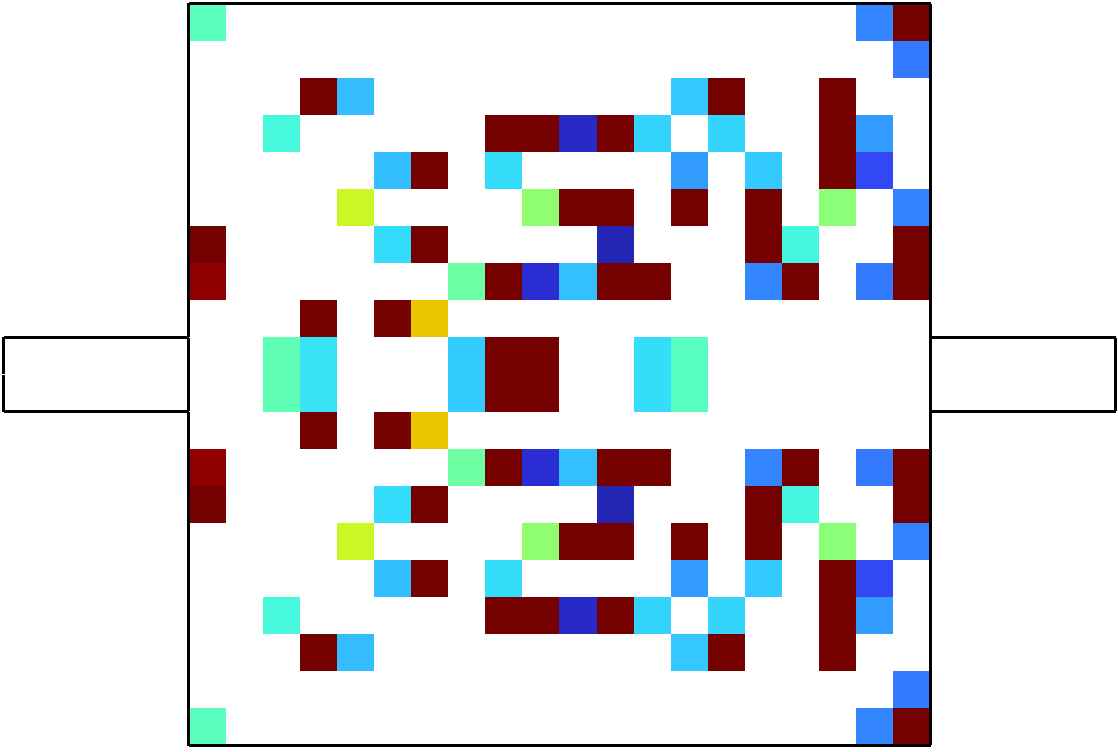}
        \caption{}
        \label{fig:dist_1Pa}
    \end{subfigure}\hfill
    \begin{subfigure}[b]{0.3\textwidth}
        \centering
        \includegraphics[height=0.7\linewidth]{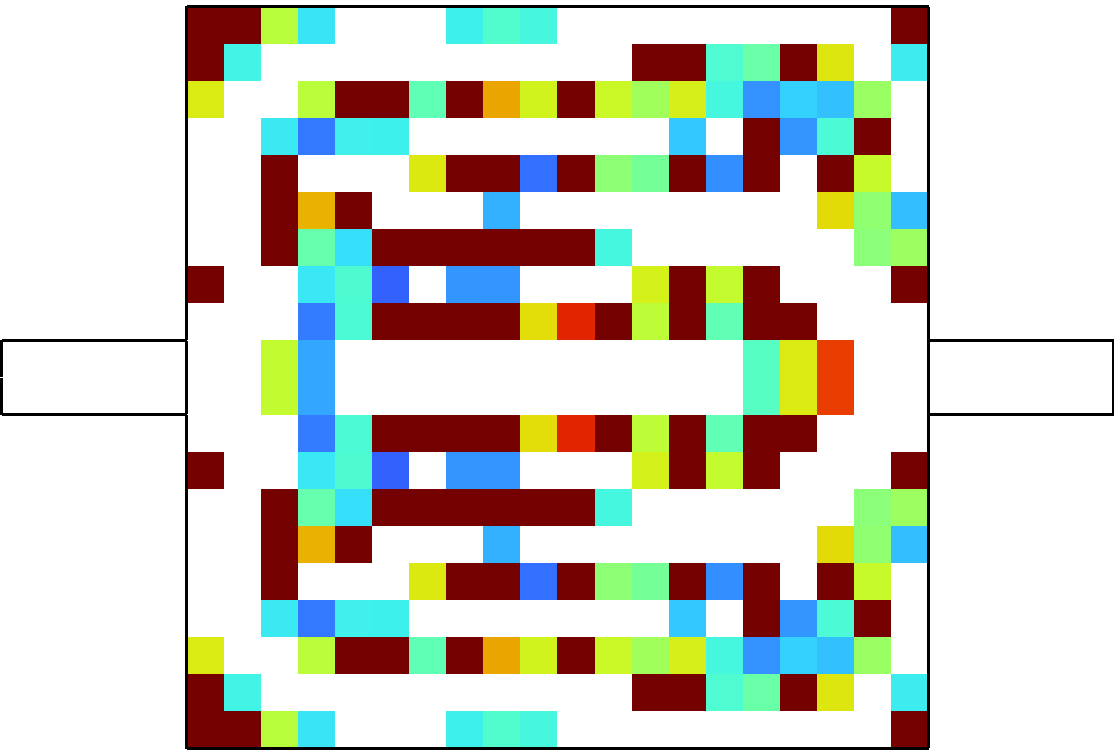}
        \caption{}
        \label{fig:dist_10Pa}
    \end{subfigure}\hfill
    \begin{subfigure}[b]{0.3\textwidth}
        \centering
        \includegraphics[height=0.7\linewidth]{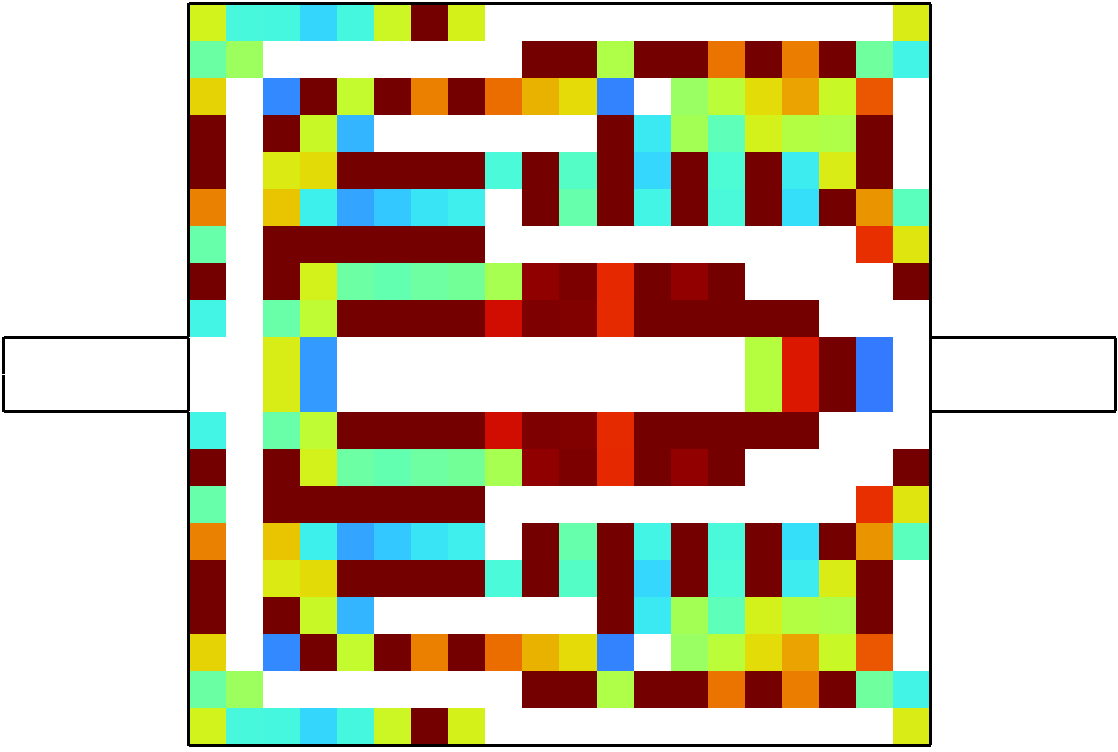}
        \caption{}
        \label{fig:dist_50Pa}
    \end{subfigure}
    
    \vspace{0.5em}
    \begin{subfigure}[b]{0.5\textwidth}
        \centering
        
        \includegraphics[width=\linewidth]{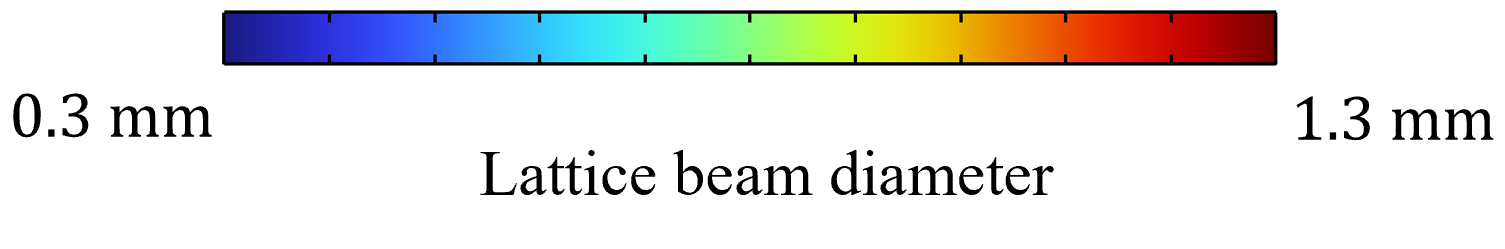}

    \end{subfigure}
    
    \vspace{0.5em}
    \begin{subfigure}[b]{0.3\textwidth}
        \centering
        \includegraphics[height=1.6\linewidth]{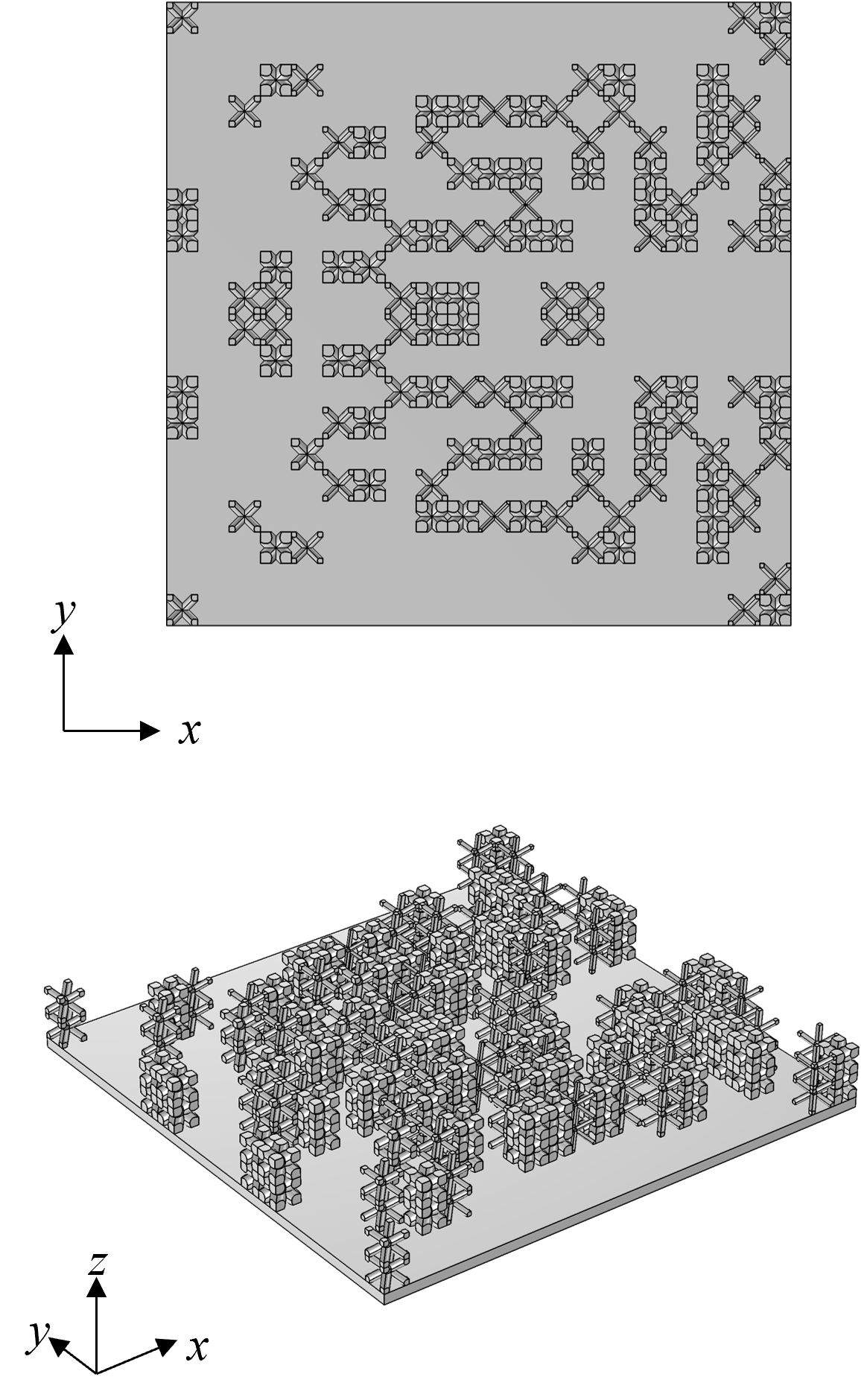}
        \caption{}
        \label{fig:FS_1Pa}
    \end{subfigure}\hfill
    \begin{subfigure}[b]{0.3\textwidth}
        \centering
        \includegraphics[height=1.6\linewidth]{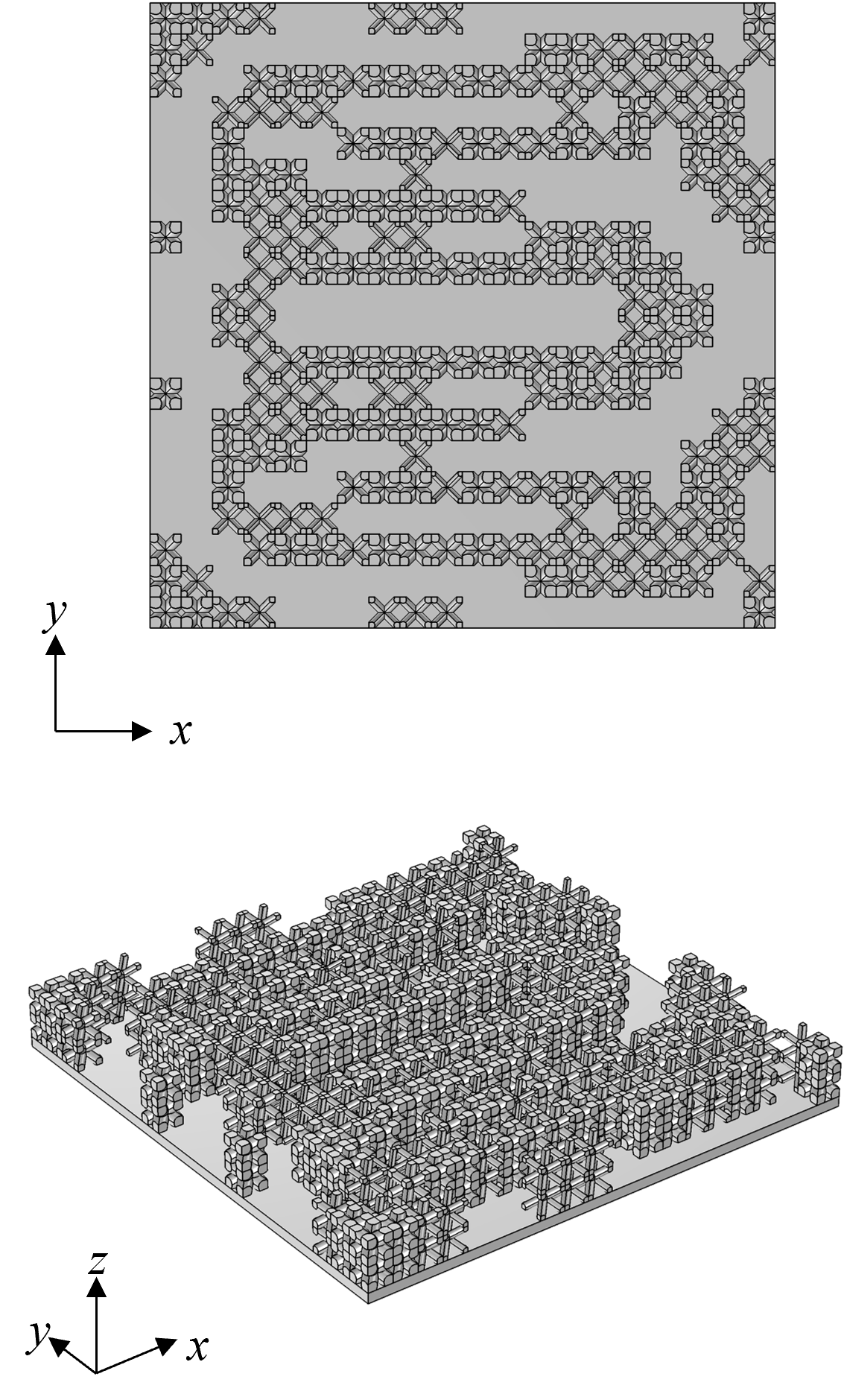}
        \caption{}
        \label{fig:FS_10Pa}
    \end{subfigure}\hfill
    \begin{subfigure}[b]{0.3\textwidth}
        \centering
        \includegraphics[height=1.6\linewidth]{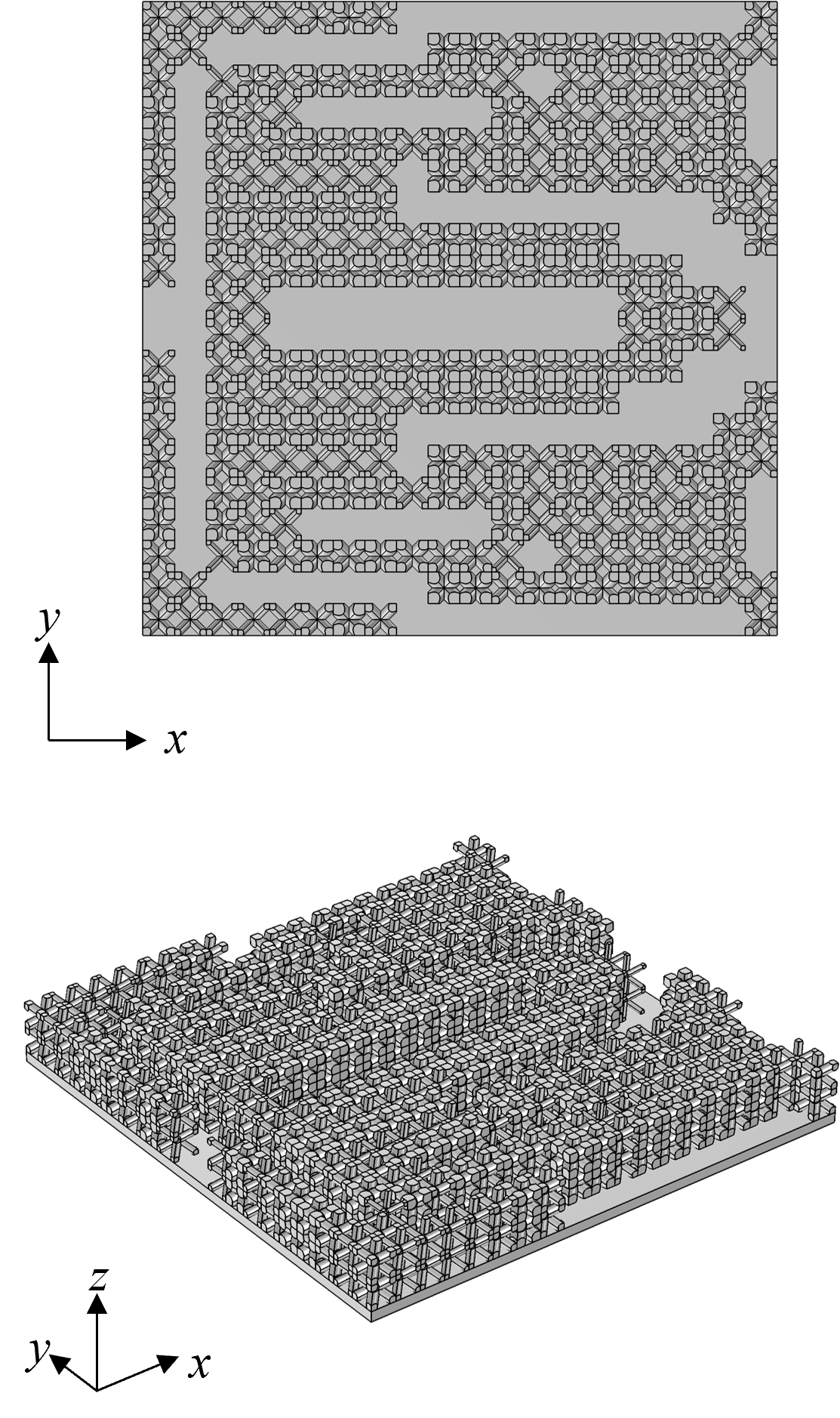}
        \caption{}
        \label{fig:FS_50Pa}
    \end{subfigure}

    \caption{Optimized distribution: (a) $P_\mathrm{in}$=1 Pa, (b) $P_\mathrm{in}$=10 Pa, (c) $P_\mathrm{in}$=50 Pa and corresponding full-scale geometries: (d) $P_\mathrm{in}$=1 Pa, (e) $P_\mathrm{in}$=10 Pa, (f) $P_\mathrm{in}$=50 Pa. }
    \label{fig:Optimal_shape}
\end{figure}

The optimized material distributions and corresponding full-scale geometries are presented in \figref{Optimal_shape}. 
The resulting shapes varied depending on the applied pressure drop: as the inlet pressure increased, regions 
with higher relative density expanded. 
This trend reflects the trade-off between enhanced heat transfer and increased pressure loss, consistent with 
previous studies on microchannel heat sinks \cite{Yanetal}. 
\subsection{Evaluation of optimized designs via full-scale analysis}
\label{subsec3.4}
\begin{figure}[htbp]
    \centering
    \begin{subfigure}[b]{0.32\textwidth}
        \includegraphics[height=1.4\textwidth]{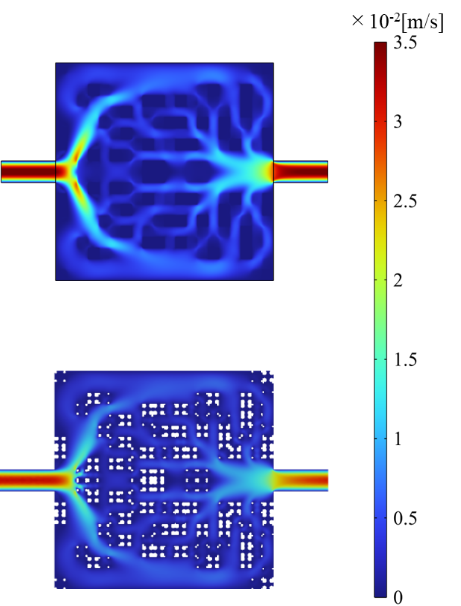}
        \caption{}
        \label{fig:velo_comp_1Pa}
    \end{subfigure}
    \begin{subfigure}[b]{0.32\textwidth}
        \includegraphics[height=1.4\textwidth]{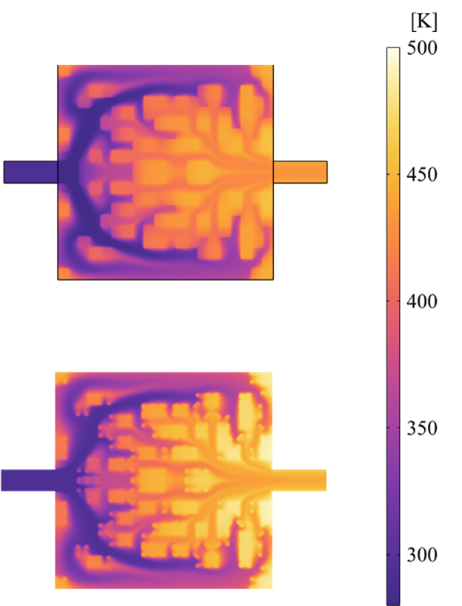}
        \caption{}
        \label{fT_comp_1Pa}
    \end{subfigure}
    \begin{subfigure}[b]{0.32\textwidth}
        \includegraphics[height=1.4\textwidth]{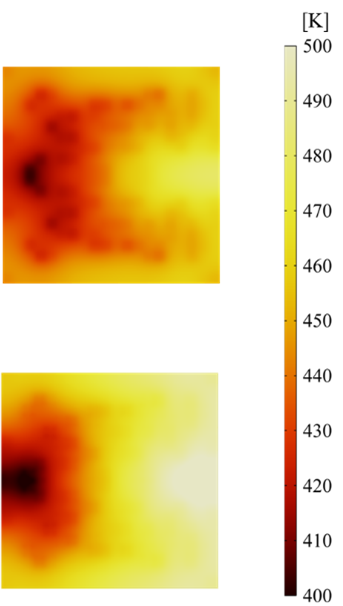}
        \caption{}
        \label{bT_comp_1Pa}
    \end{subfigure}
    \caption{Comparison between two models under $P_\mathrm{in}$=1 Pa condition: (a) Velocity field of thermal-fluid layer, (b) Temperature field of thermal-fluid layer, (c) Temperature field of base-solid layer. The upper subfigures of (a)-(c) are those calculated by equivalent model based on two-layer DF model and the lower subfigures are those of full-scale model. }
    \label{fig:1Pa_field_comp}
  \end{figure}

The optimized geometries are further evaluated using detailed full-scale analysis(full-scale model) under symmetry-reduced domains. 
%
\textbf{Figs.~\ref{fig:1Pa_field_comp}}--\textbf{\ref{fig:50Pa_field_comp}} compare the velocity and temperature fields obtained from the two-layer DF model and the full-scale analysis for the three pressure conditions. 
The two-layer DF model is obtained by integrating the governing equations for incompressible thermal fluid flow along the thickness direction. 
Therefore, the quantities such as velocity are expressed in terms of thickness-integrated values. 
Consequently, the correction factors (Eqs.~\eqref{eq:repair_velocity} and~\eqref{eq:repair_temp}) must be applied when comparing two-layer DF model and 3D full-scale model. 
  \begin{equation}
    \centering
    \bm{V}_0=\frac{3}{2} {\bm{\bar{v}}}
    \label{eq:repair_velocity}
  \end{equation}

  \begin{equation}
    \centering
    T(x,y,H_\mathrm{t})=-\frac{39}{416} \left(T_\mathrm{b0}-\frac{q}{k_\mathrm{b}} H_\mathrm{b} \right)+\frac{455}{416}T_0
    \label{eq:repair_temp}
  \end{equation}
Since the inlet and outlet regions, which are outside the design domain, are not heated from the bottom surface, it is assumed that the fluid is sufficiently mixed in these areas. 
Therefore, the temperature distribution in the z-direction is neglected, and the bulk temperature ($T_0$) is directly plotted. 
On the other hand, velocity field and temperature field of central section in each phase are plotted in full-scale analysis model.

In general, the DF model predicted lower temperature distributions compared with the full-scale model. 
This discrepancy is attributed to the simplified assumption of smooth boundary layers in the DF formulation. 
%

Quantitative comparisons are presented in \tabref{Error_equvsFS}. Across all pressure conditions, the DF model slightly overestimated velocities and Nusselt numbers, with errors of approximately 7--23\%. 
Despite these deviations, the results confirm that the DF model provides sufficiently accurate predictions to guide optimization.

A cross-check is also performed to evaluate the robustness of the optimization. Optimized geometries obtained under one pressure condition are re-analyzed at different pressures using both two-layer DF model and the full-scale model. 
The results (\tabref{crosscheck}) showed that each design achieved the highest $Nu_\mathrm{obj}$
under the condition for which it is optimized, demonstrating the validity of the optimization strategy.

  \begin{figure}[htbp]
    \centering
    \begin{subfigure}[b]{0.32\textwidth}
        \includegraphics[height=1.4\textwidth]{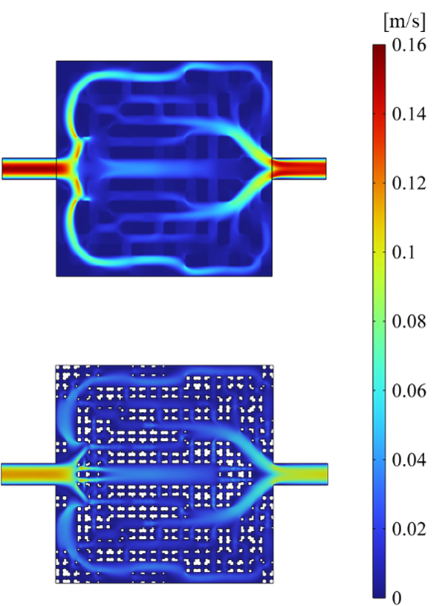}
        \caption{}
        \label{fig:velo_comp_10Pa}
    \end{subfigure}
    \begin{subfigure}[b]{0.32\textwidth}
        \includegraphics[height=1.4\textwidth]{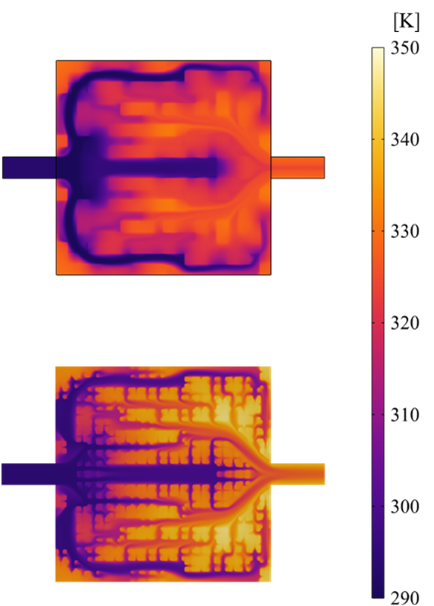}
        \caption{}
        \label{fT_comp_10Pa}
    \end{subfigure}
    \begin{subfigure}[b]{0.32\textwidth}
        \includegraphics[height=1.4\textwidth]{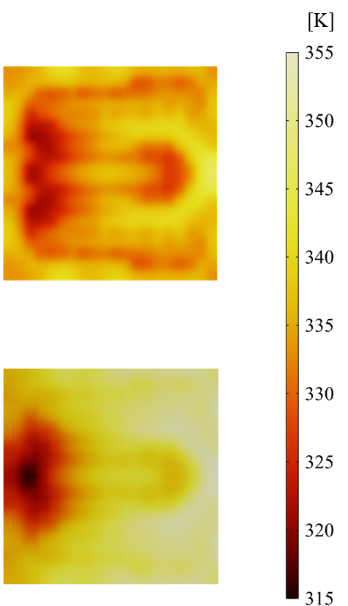}
        \caption{}
        \label{bT_comp_10Pa}
    \end{subfigure}
    \caption{Comparison between two models under $P_\mathrm{in}$=10 Pa condition: (a) Velocity field of thermal-fluid layer, (b) Temperature field of thermal-fluid layer, (c) Temperature field of base-solid layer. The upper subfigures of (a)-(c) are those calculated by equivalent model based on two-layer DF model and the lower subfigures are those of full-scale model. }
    \label{fig:10Pa_field_comp}
  \end{figure}

  \begin{figure}[htbp]
    \centering
    \begin{subfigure}[b]{0.32\textwidth}
        \includegraphics[height=1.4\textwidth]{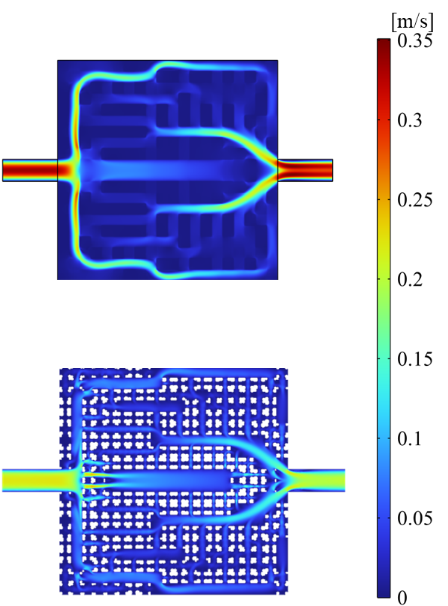}
        \caption{}
        \label{fig:velo_comp_50Pa}
    \end{subfigure}
    \begin{subfigure}[b]{0.32\textwidth}
        \includegraphics[height=1.4\textwidth]{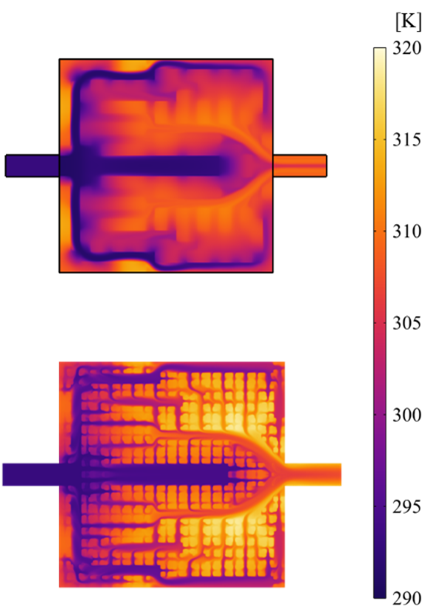}
        \caption{}
        \label{fT_comp_50Pa}
    \end{subfigure}
    \begin{subfigure}[b]{0.32\textwidth}
        \includegraphics[height=1.4\textwidth]{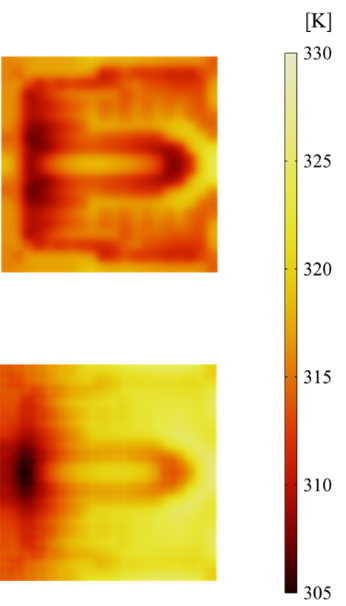}
        \caption{}
        \label{bT_comp_50Pa}
    \end{subfigure}
    \caption{Comparison between two models under $P_\mathrm{in}$=50 Pa condition: (a) Velocity field of thermal-fluid layer, (b) Temperature field of thermal-fluid layer, (c) Temperature field of base-solid layer. The upper subfigures of (a)-(c) are those calculated by equivalent model based on two-layer DF model and the lower subfigures are those of full-scale model. }
    \label{fig:50Pa_field_comp}
  \end{figure}
  \begin{table}[htbp]
    \centering
    \caption{Error between two-layer DF model and full-scale model. }
    \begin{tabular}{cccc}
    \hline
    \multirow{2}{*}{$P_{\mathrm{in}} \mathrm{[Pa]}$} & \multicolumn{3}{c}{$u_{\mathrm{in}} \mathrm{[m/s]}$}                                                   \\ \cline{2-4} 
                                  & \multicolumn{1}{c}{Two-layer DF} & \multicolumn{1}{c}{Full-scale} & Error [\%] \\ \hline
    1                             & \multicolumn{1}{c}{$1.70\times10^{-2}$}    & \multicolumn{1}{c}{$1.51\times10^{-2}$}  & 13.1           \\
    10                            & \multicolumn{1}{c}{$7.22\times10^{-2}$}    & \multicolumn{1}{c}{$6.21\times10^{-2}$}  & 16.3           \\
    50                            & \multicolumn{1}{c}{0.156}        & \multicolumn{1}{c}{0.138}      & 12.6           \\ \hline
    \multirow{2}{*}{$P_{\mathrm{in}} \mathrm{[Pa]}$} & \multicolumn{3}{c}{$Nu_\mathrm{max}$}                                                           \\ \cline{2-4} 
                                  & \multicolumn{1}{c}{Two-layer DF}             & \multicolumn{1}{c}{Full-scale} & Error [\%] \\ \hline
    1                             & \multicolumn{1}{c}{4.44}         & \multicolumn{1}{c}{3.89}       & 14.3        \\   
    10                            & \multicolumn{1}{c}{15.4}         & \multicolumn{1}{c}{13.5}       & 14.1         \\  
    50                            & \multicolumn{1}{c}{26.6}         & \multicolumn{1}{c}{24.8}       & 7.27            \\ \hline
    \multirow{2}{*}{$P_{\mathrm{in}} \mathrm{[Pa]}$} & \multicolumn{3}{c}{$Nu_\mathrm{obj}$}                                                           \\ \cline{2-4} 
                                  & \multicolumn{1}{c}{Two-layer DF} & \multicolumn{1}{c}{Full-scale} & Error [\%] \\ \hline
    1                             & \multicolumn{1}{c}{5.32}         & \multicolumn{1}{c}{4.59}       & 16.0        \\   
    10                            & \multicolumn{1}{c}{19.8}         & \multicolumn{1}{c}{15.9}       & 24.8        \\  
    50                            & \multicolumn{1}{c}{37.1}         & \multicolumn{1}{c}{30.4}       & 22.0           \\ \hline
    \multirow{2}{*}{$P_{\mathrm{in}} \mathrm{[Pa]}$} & \multicolumn{3}{c}{$Nu_\mathrm{ave}$}                                                           \\ \cline{2-4} 
                                  & \multicolumn{1}{c}{Two-layer DF} & \multicolumn{1}{c}{Full-scale} & Error [\%] \\ \hline
    1                             & \multicolumn{1}{c}{5.54}         & \multicolumn{1}{c}{4.93}       & 12.3        \\   
    10                            & \multicolumn{1}{c}{20.5}         & \multicolumn{1}{c}{17.1}       & 20.0        \\   
    50                            & \multicolumn{1}{c}{38.6}         & \multicolumn{1}{c}{32.8}       & 19.4           \\ \hline
    \end{tabular}
    \label{tab:Error_equvsFS}
  \end{table}

  \begin{table}[htbp]
    \centering
    \caption{Crosscheck of $Nu_{\mathrm{obj}}$: (a) two-layer DF model, (b) full-scale model. }
    \begin{subtable}[t]{0.45\textwidth}
      \centering
      \caption{}
      \begin{tabular}{cccc}
      \hline
      \multirow{2}{*}{Analysis} & \multicolumn{3}{c}{Optimization}                            \\ 
                                & \multicolumn{1}{c}{1~Pa}  & \multicolumn{1}{c}{10~Pa} & \multicolumn{1}{c}{50~Pa} \\ \hline
      1~Pa                       & \multicolumn{1}{c}{\textcolor{blue}{5.32}} & \multicolumn{1}{c}{3.66} & \multicolumn{1}{c}{2.05} \\ 
      10~Pa                      & \multicolumn{1}{c}{10.6} & \multicolumn{1}{c}{\textcolor{blue}{19.8}} & \multicolumn{1}{c}{15.3} \\ 
      50~Pa                      & \multicolumn{1}{c}{12.0} & \multicolumn{1}{c}{30.9} & \multicolumn{1}{c}{\textcolor{blue}{37.1}} \\ \hline
      \end{tabular}
      \label{tab:DF_crosscheck}
    \end{subtable}

    \vspace{1em} 
    
    \begin{subtable}[t]{0.45\textwidth}
      \centering
      \caption{}
      \begin{tabular}{cccc}
      \hline
      \multirow{2}{*}{Analysis} & \multicolumn{3}{c}{Optimization}                            \\ 
                                & \multicolumn{1}{c}{1~Pa}  & \multicolumn{1}{c}{10~Pa} & \multicolumn{1}{c}{50~Pa} \\ \hline
      1~Pa                       & \multicolumn{1}{c}{\textcolor{blue}{4.59}} & \multicolumn{1}{c}{3.37} & \multicolumn{1}{c}{1.94} \\ 
      10~Pa                      & \multicolumn{1}{c}{12.5} & \multicolumn{1}{c}{\textcolor{blue}{15.9}} & \multicolumn{1}{c}{12.9} \\ 
      50~Pa                      & \multicolumn{1}{c}{20.4} & \multicolumn{1}{c}{29.0} & \multicolumn{1}{c}{\textcolor{blue}{30.4}} \\ \hline
      \end{tabular}
      \label{tab:fS_crosscheck}
    \end{subtable}
    \label{tab:crosscheck}
  \end{table}

\subsection{Introducing reference fins}
\label{subsec3.5}
Here, the following three types of reference fins are prepared (Fig. 17) to benchmark performance:
\begin{itemize}
\item 
A uniform BCC lattice with beam diameter $d=0.3$~mm, corresponding to the initial design.
\item
A non-uniform lattice without voids, optimized following the method of Takezawa et al. [32], with strut diameters ranging from 0.3-1.3 mm.
\item A conventional plate-fin heat sink, selected from a parametric study to maximize $Nu_\mathrm{max}$
under a pressure drop of 50 Pa.
\end{itemize} 
%
A parametric study is conducted on the several plate-fin with different fin thickness, number of fins, and inlet clearance, 
while the outlet clearance of all plate fins is fixed at $L$ ($2.5 \: \mathrm{mm}$) to simplify the parameter study.

\begin{figure}[htbp]
    \centering
    %
    \begin{subfigure}[b]{0.3\textwidth}
      \centering
      \includegraphics[height=1.6\linewidth]{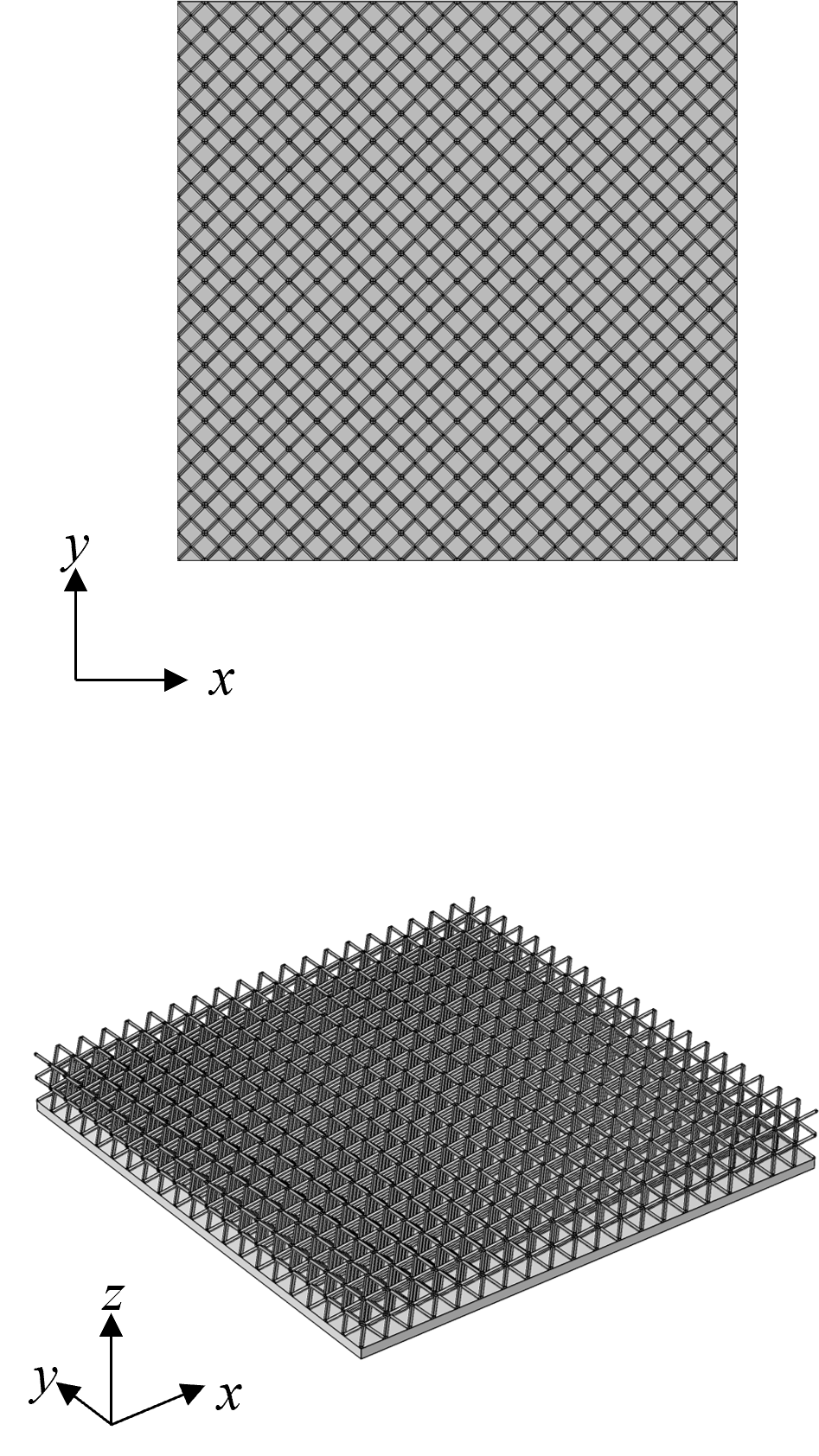}
      \caption{}
      \label{fig:UniformBCC}
    \end{subfigure}
    \hfill
    %
    \begin{subfigure}[b]{0.3\textwidth}
      \centering
      \includegraphics[height=1.6\linewidth]{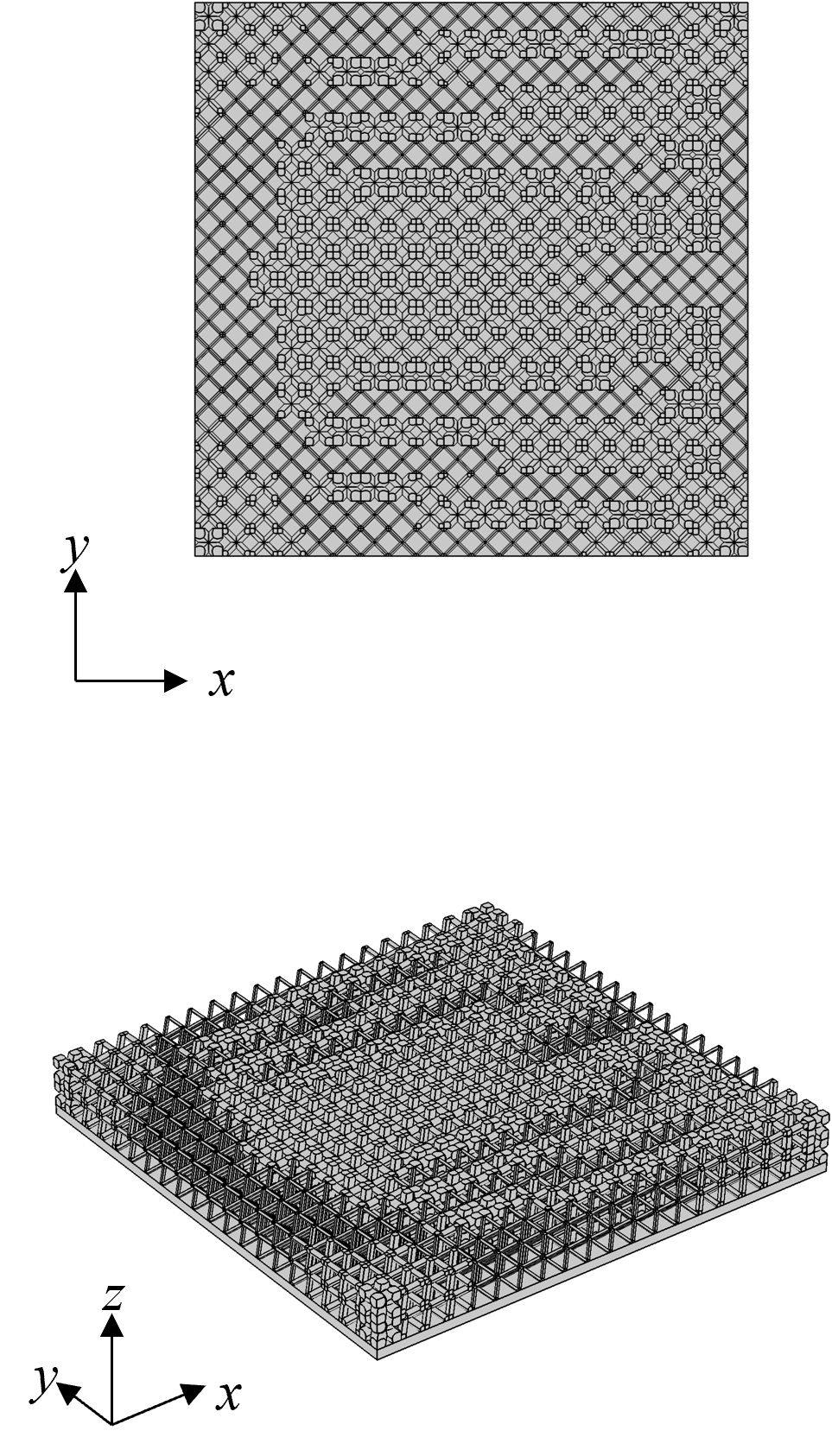}
      \caption{}
      \label{fig:LBTO_BCC}
    \end{subfigure}
    \hfill
    \begin{subfigure}[b]{0.3\textwidth}
      \centering
      \includegraphics[height=1.6\linewidth]{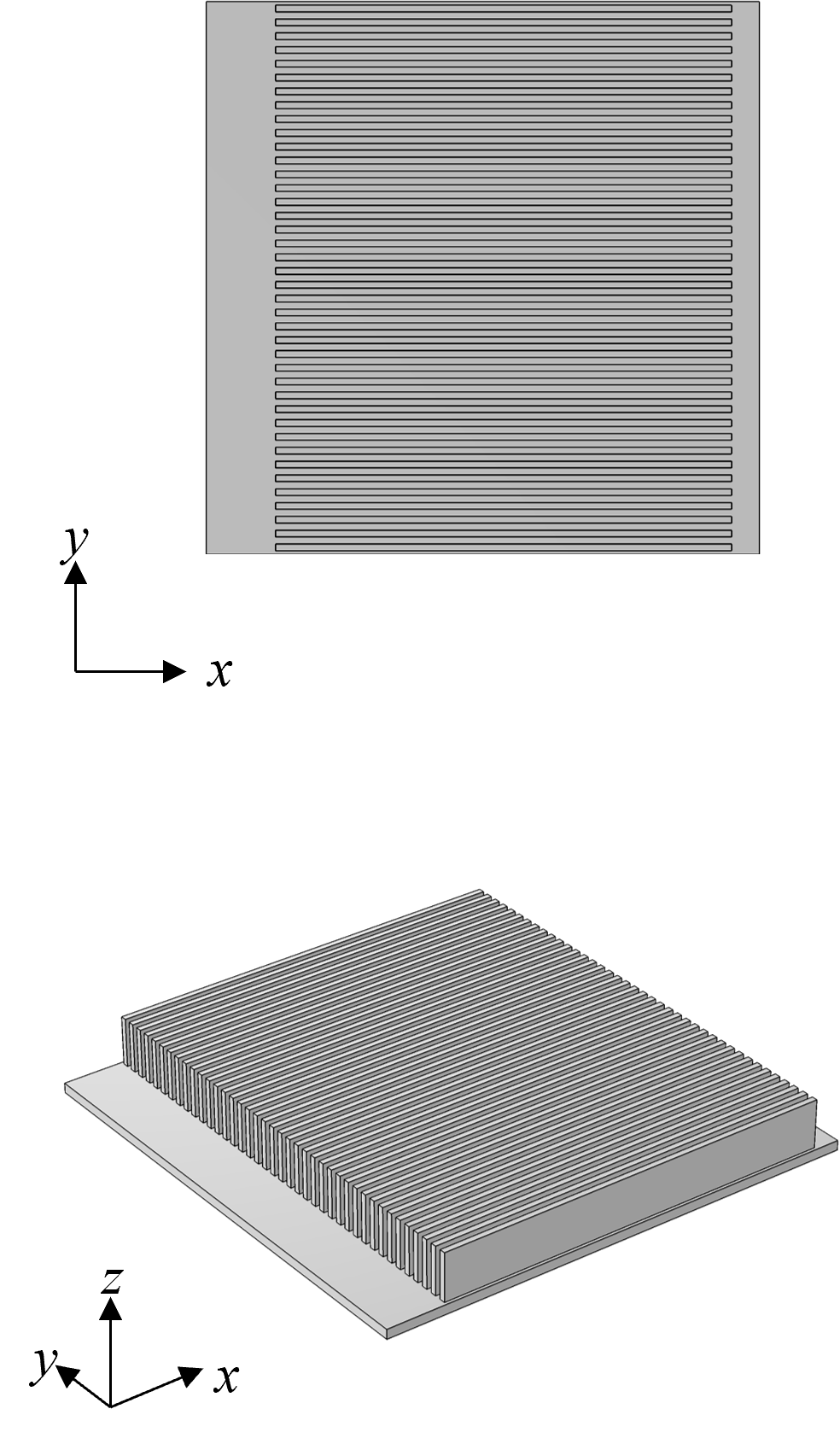}
      \caption{}
      \label{fig:Plate}
    \end{subfigure}
    \caption{Reference fins used in the performance evaluation: (a) Uniform BCC, (b) Optimized BCC without void, (c) Plate-fin.}
    \label{fig:Reference}
  \end{figure}

In order to ensure the validity of the comparison, 
FE analyses are also conducted under fixed pressure condition of 50 Pa for each configuration, and the design with the highest $Nu_\mathrm{max}$ is identified. 
As a result, the optimized configuration is found to have a plate thickness of 0.606 mm (solid fraction = 0.4), 40 fins, and a inlet clearance of $2.5L$ ($6.25 \: \mathrm{mm}$)
The reference plate-fin is illustrated in \figref{Plate}. 
\subsection{Performance evaluation}
\label{subsec3.6}
All designs are evaluated using full-scale FE simulations under a uniform inlet pressure of 50 Pa.
The uniform distributed BCC lattice, the non-void optimized BCC lattice, optimized voided BCC lattice and the plate-fin are compared using $Nu_{\mathrm{max}}$ and $Nu_{\mathrm{ave}}$.

\begin{figure}[htbp]
    \centering
    \begin{subfigure}{\textwidth}
      \centering
      \includegraphics[width=0.9\linewidth]{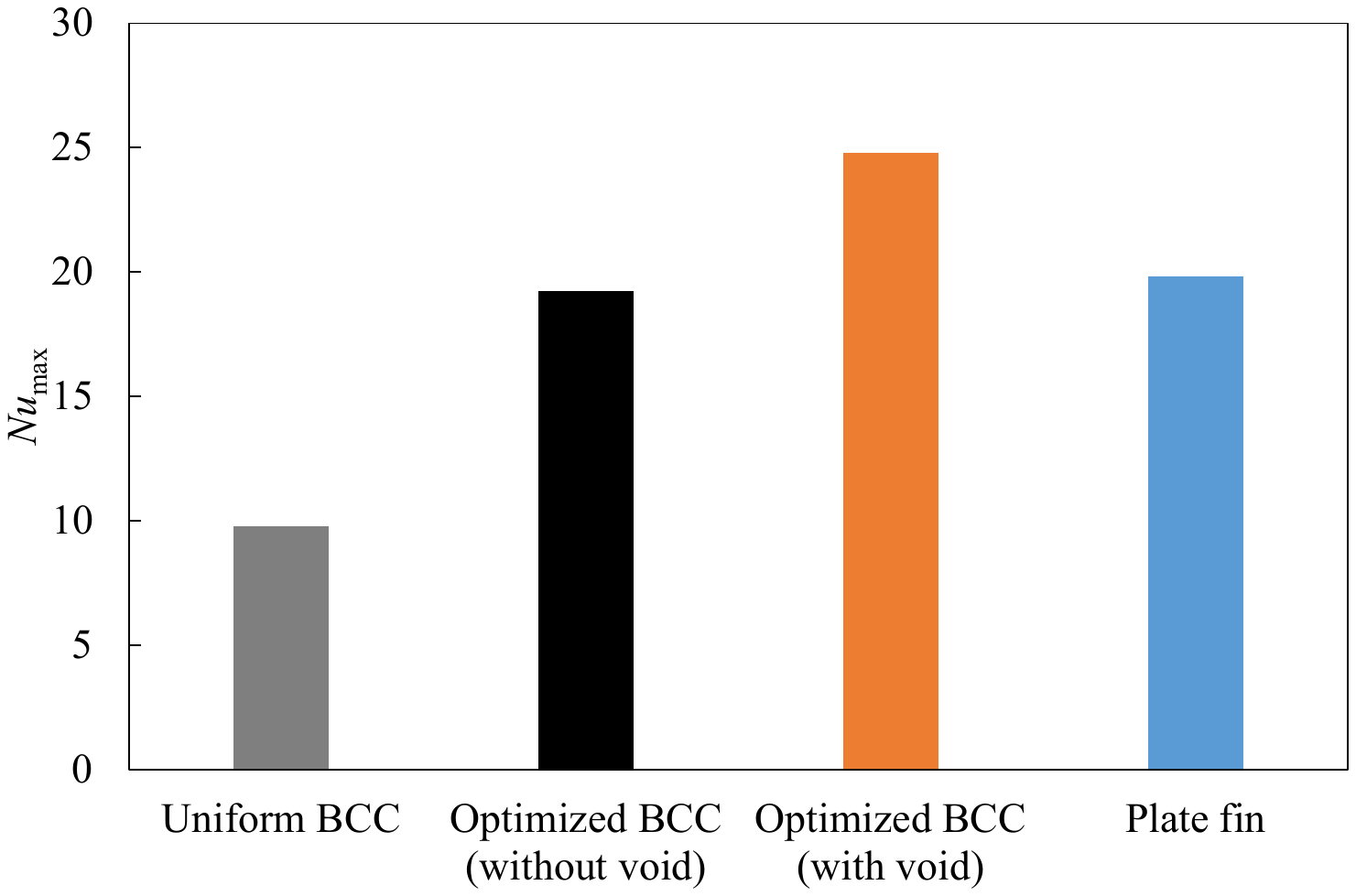}
      \caption{}
      \label{fig:comp_Numax}
    \end{subfigure}
    \begin{subfigure}{\textwidth}
      \centering
      \includegraphics[width=0.9\linewidth]{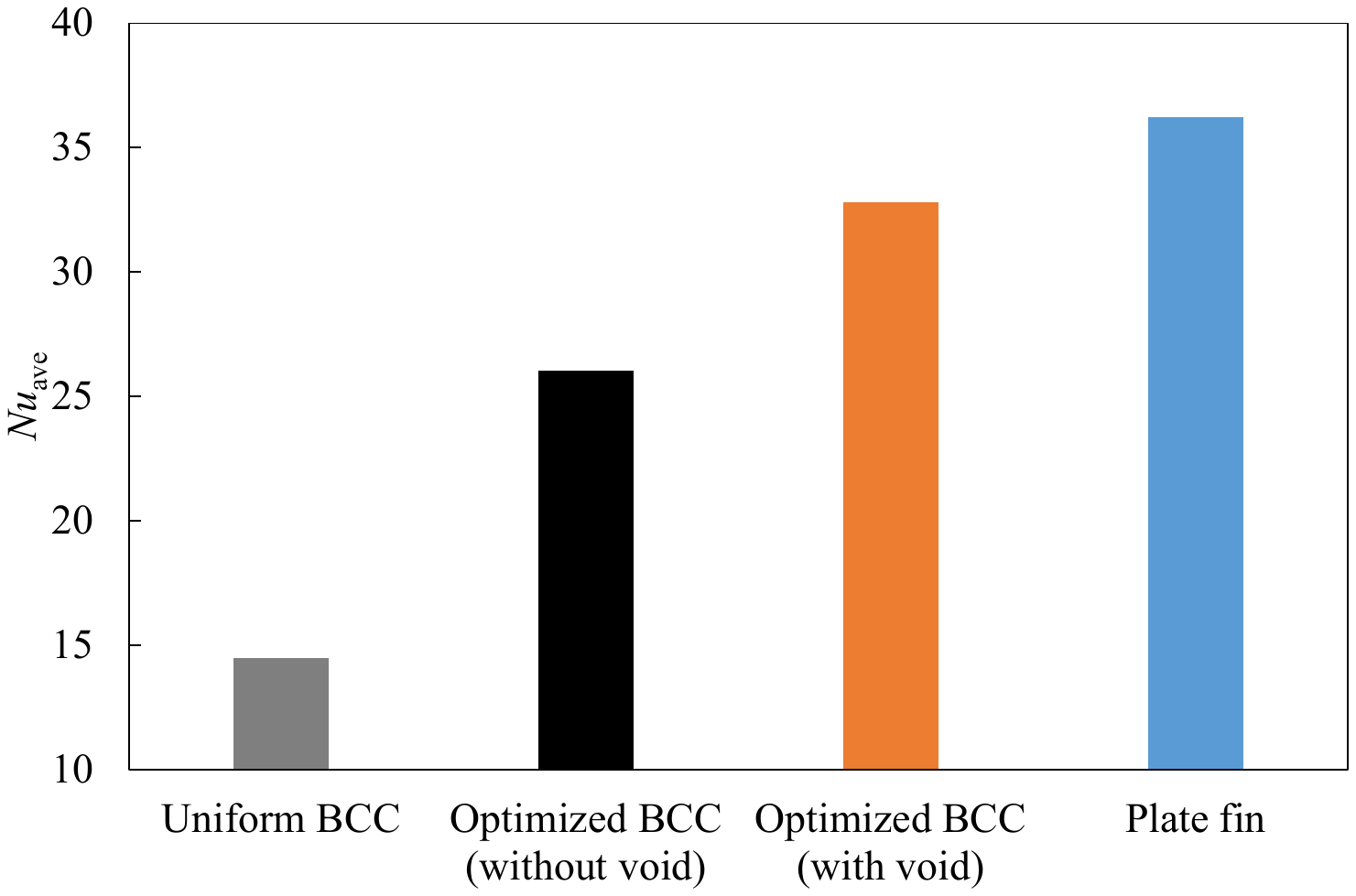}
      \caption{}
      \label{fig:comp_Nuave}
    \end{subfigure}
    \caption{Performance evaluation through comparison with reference fins: (a) $Nu_{\mathrm{max}}$, (b) $Nu_{\mathrm{ave}}$. }
  \label{fig:comparison}
\end{figure}

\begin{figure}[htbp]
    \centering
    \includegraphics[width=0.8\linewidth]{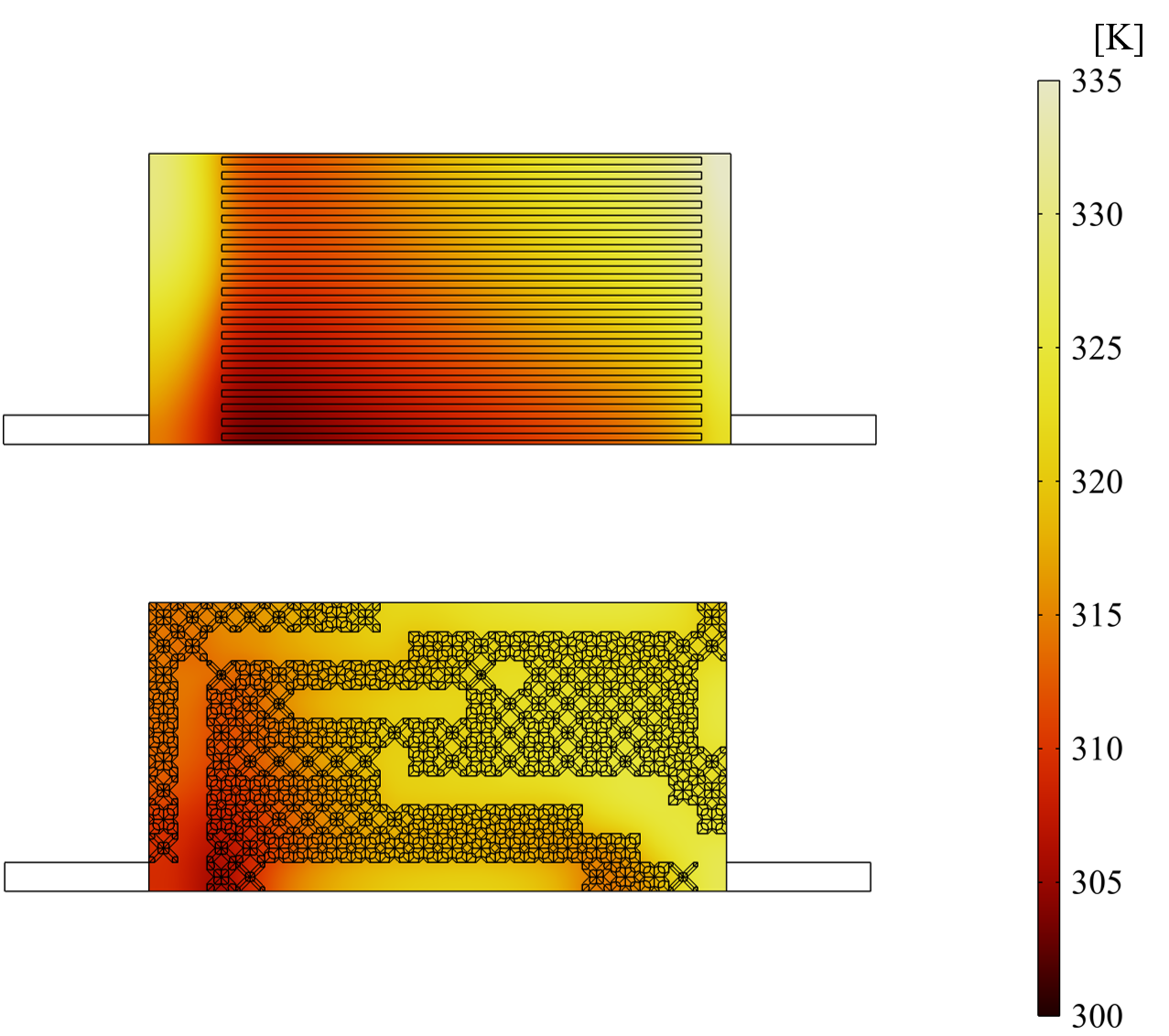}
    \caption{Comparison of base temperature between plate-fin and optimized BCC fin. }
    \label{fig:comp_temp_platevsvlbto_dark}
  \end{figure}

As shown in \figref{comparison}, the optimized lattice structures—both with and without voids—outperformed the uniform lattice. 
While introducing void regions further improved performance. 
%
Compared with the reference plate-fin with $Nu_\mathrm{max}$ of 19.8, the optimized voided lattice fin achieved 24.8, indicating 24.2\% enhancement. 
However, its $Nu_\mathrm{ave}$ (32.8) is approximately 8\% lower than that of the reference plate-fin (36.2).
The bottom-surface temperature distributions (\figref{comp_temp_platevsvlbto_dark}) explain this discrepancy. In the plate-fin design, localized jet impingement produces regions of intense cooling, yielding high $Nu_\mathrm{ave}$ but uneven temperature dispersion. 
In contrast, the optimized lattice achieves a more uniform temperature distribution across the base plate, 
indicating superior thermal management despite a slightly lower $Nu_\mathrm{ave}$.

\subsection{Effect of lattice cell miniaturization}
\label{subsec3.7}
In the above discussion, the unit size of the BCC lattice is fixed at 2.5 mm. 
However, this reduces the number of design variables and limits optimization resolution, 
thereby, restricting the performance improvement. 
Cell miniaturization can be the powerful solution to address this issue.
Furthermore, performance enhancement is expected due to the increased surface area result from structural refinement. 
However, smaller unit cells also introduce challenges: (i) higher pressure drops due to increased pore density, 
and (ii) manufacturability limits, since beams with too small diameter are difficult to fabricate using powder 
bed fusion (PBF) with high-conductivity metals such as aluminum alloys.

\begin{figure}[htbp]
      \centering
      \begin{subfigure}{0.6\textwidth}
        \centering
        \includegraphics[width=0.9\linewidth]{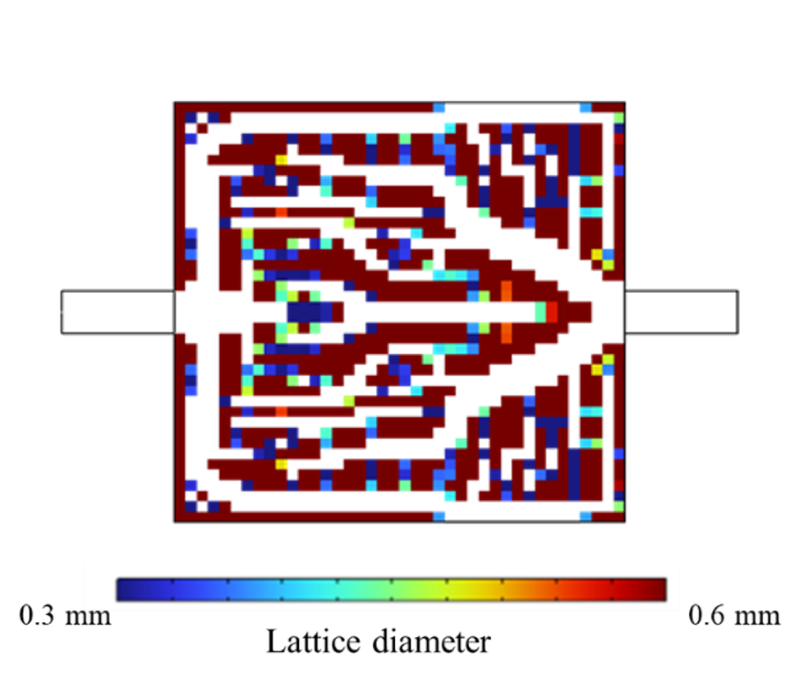}
        \caption{}
        \label{fig:1.25_equi}
      \end{subfigure}
      
      \begin{subfigure}{0.6\textwidth}
        \centering
        \includegraphics[width=0.75\linewidth]{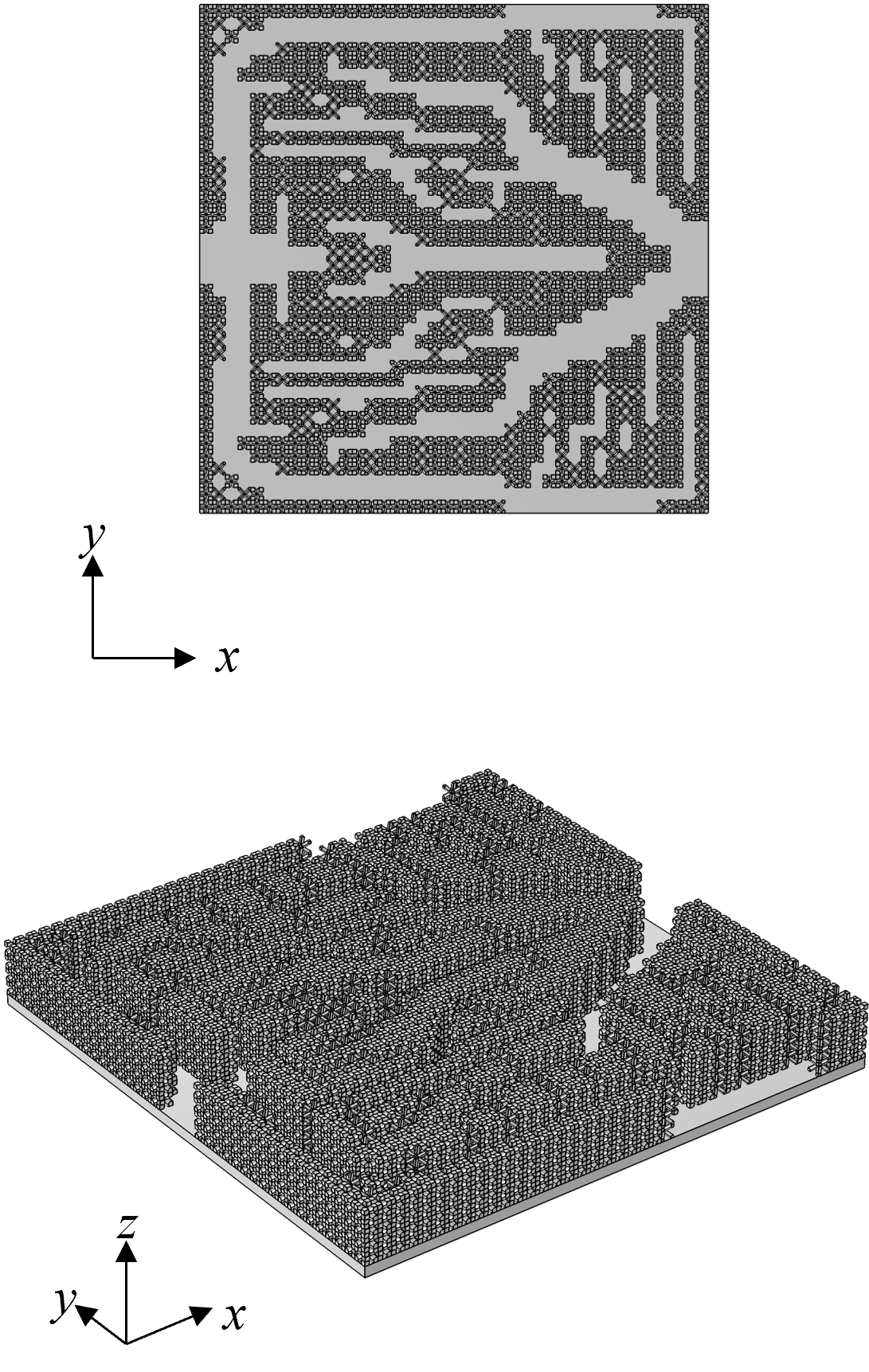}
        \caption{}
        \label{fig:1.25_FS}
      \end{subfigure}
      \caption{Optimized material distribution for 1.25 mm unit cell size ($P_\mathrm{in}$=50 Pa): (a) Distribution of non-uniform lattice, (b) Corresponding full-scale geometry. }
    \label{fig:1.25_opt}
  \end{figure}

\begin{figure}[htbp]
      \centering
      \begin{subfigure}{\textwidth}
        \centering
        \includegraphics[width=0.9\linewidth]{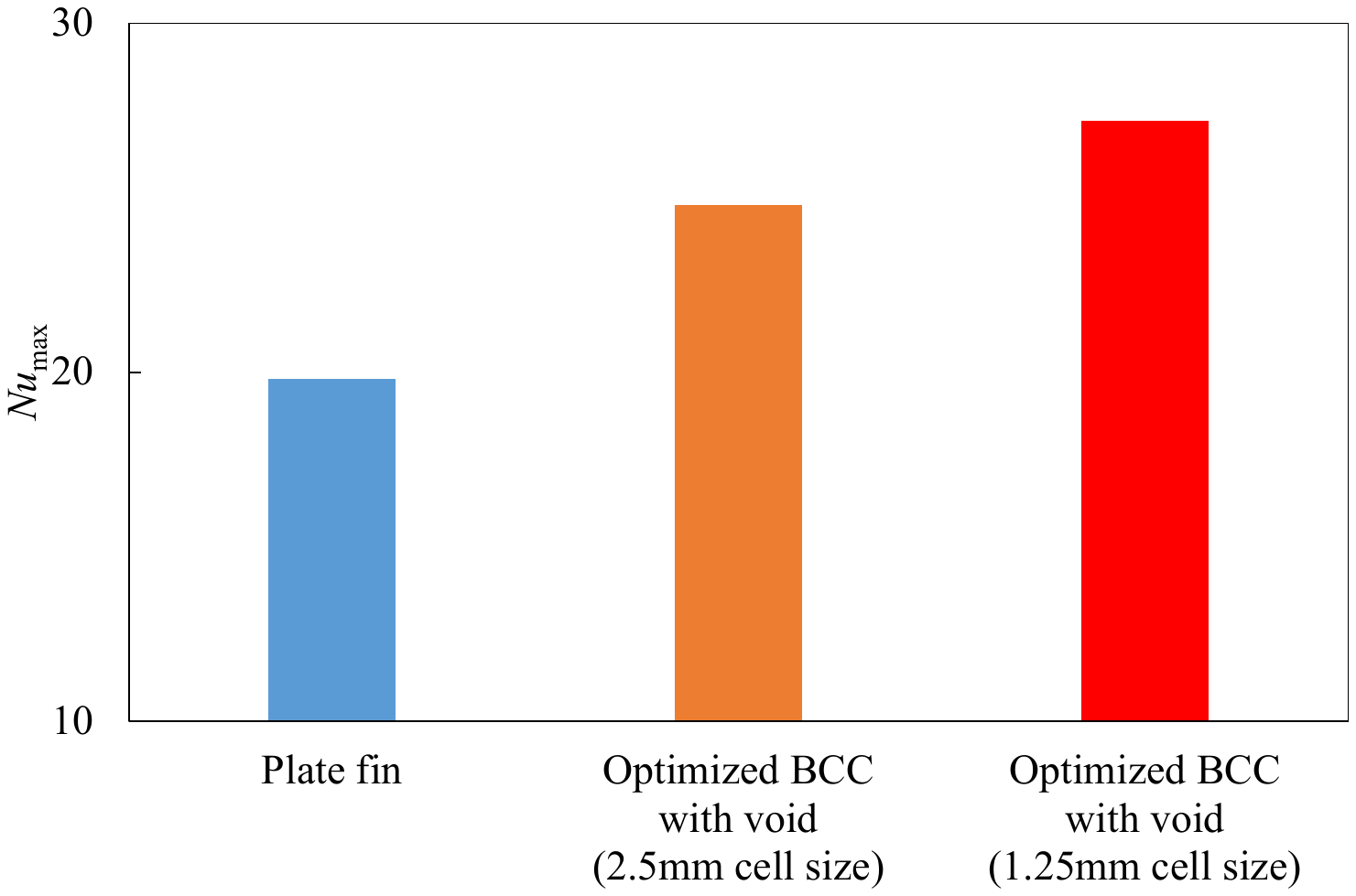}
        \caption{}
        \label{fig:1.25mm_Numax}
      \end{subfigure}
      \begin{subfigure}{\textwidth}
        \centering
        \includegraphics[width=0.9\linewidth]{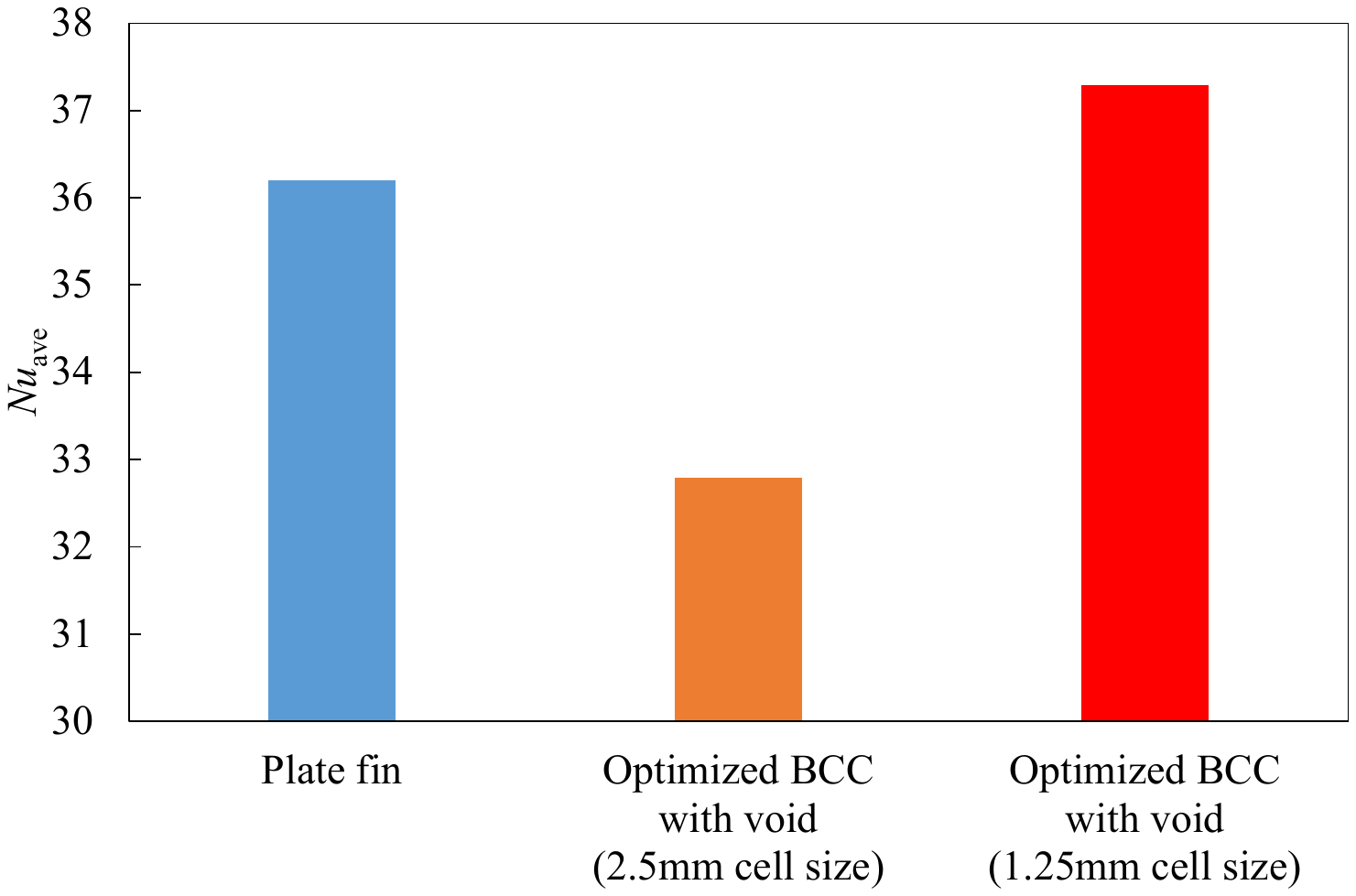}
        \caption{}
        \label{fig:1.25mm_Nuave}
      \end{subfigure}
      \caption{Performance comparison of optimized miniaturized cell with others: (a) $Nu_{\mathrm{max}}$, (b) $Nu_{\mathrm{ave}}$}
      \label{fig:1.25mmcomp}
    \end{figure}

To investigate this trade-off, the unit cell size is reduced to 1.25 mm, with beam diameters constrained between 0.3 and 0.6 mm. 
The optimized distribution for the 50 Pa case is shown in \figref{1.25_opt}, and the performance comparison is conducted based on full-scale model.
The result presented in \figref{1.25mmcomp}. 
The miniaturized lattice achieved further improvements in both $Nu_\mathrm{max}$ and $Nu_\mathrm{ave}$, 
demonstrating that cell miniaturization, despite manufacturing challenges, can yield substantial gains in thermal performance.

\section{Conclusion}
\label{sec4}
In this study, a multi-material topology optimization framework is applied to design heat sinks incorporating both void regions and graded lattice structures. 
By explicitly embedding voids within heterogeneous microstructures, the proposed approach successfully mitigates the pressure drop issue that has limited the application of porous media in thermal management.
Even under identical pressure-drop conditions, the optimized designs achieved performance improvements of approximately 20--30\% 
compared with conventional fin-type heat sinks. 
Despite certain limitations, particularly the reduced fidelity of the equivalent model, the results 
clearly demonstrate the potential of combining graded lattices with void regions to realize next-generation cooling devices with enhanced 
thermal performance and energy efficiency.
For the further advance in this study, we will conduct experimental validation to evaluate the performance of the proposed design under practical operating conditions. 
Such investigations are expected to provide deeper insights into both the accuracy and the applicability of the present approach.
\clearpage
\appendix
\section{Derivation of two-layer DF model}
\label{app1}
\subsection{Governing equation of fluid}
The governing equations of incompressible fluid can be expressed as:  
  \begin{equation}
    \centering
    \nabla \cdot \bm{v}=0
    \label{eq:continuity_eq}
  \end{equation}
  \begin{equation}
    \centering
    \rho_\mathrm{f} (\bm{v} \cdot \nabla)\bm{v}=-\nabla p +\mu_\mathrm{f} \nabla^2 \bm{v}
    \label{eq:navie_eq}
  \end{equation}
In this study, the fluid flow is assumed to be hydrodynamically developped and the design domain is significantly larger in the in plane ($x$--$y$) directions relative to its thickness ($z$--direction). 
Based on poiseuille flow theory, the velocity profile can be introduced as: 
  \begin{equation}
    \centering
    \bm{v} = \bm{V}_0(x, y) \left[ 1 - \left( \frac{z}{H_\mathrm{t}} \right)^2 \right]
    \label{eq:velocity_profile_0}
  \end{equation}
where, $\bm{V}_0$ refer to the velocity in the central plane of thermal-fluid layer. 
While, the difinition of the darcy velocity is given as: 
  \begin{equation}
    \centering
    \bm{\bar{v}}(x,y)=\frac{1}{2H_\mathrm{t}}\int_{-H_\mathrm{t}}^{H_\mathrm{t}} \bm{v} dz
    \label{eq:darcy_velocity_def}
  \end{equation}
darcy velocity $\bm{\bar{v}}$ can be calculated as:
  \begin{equation}
    \centering
    \bm{\bar{v}}=\frac{2}{3}\bm{V}_0
    \label{eq:darcy_velocity}
  \end{equation}
\par
Substitute Eq.~\eqref{eq:darcy_velocity} into \eqref{eq:velocity_profile_0}, the velocity profile along the height of thermal-fluid layer can finally be derived as follows: 
  \begin{equation}
    \centering
    \bm{v} = \frac{3}{2} \bm{\bar{v}}(x, y) \left[ 1 - \left( \frac{z}{H_\mathrm{t}} \right)^2 \right]
    \label{eq:velocity_profile}
  \end{equation}
\par
The reduced two-dimensional continuity equation and Navie-Stokes equation are obtained by averaging Eq.~\eqref{eq:continuity_eq} and \eqref{eq:navie_eq} over the domain $z\in[-H_\mathrm{t} , H_\mathrm{t}]$: 

  \begin{equation}
    \centering
    \nabla \cdot \bar{\bm{v}}=0
    \label{eq:continuity_eq_2d}
  \end{equation}

  \begin{equation}
    \centering
    \frac{6}{5}\rho_\mathrm{f} (\bar{\bm{v}} \cdot \nabla)\bar{\bm{v}}=-\nabla p +\mu_\mathrm{f} \nabla^2 \bar{\bm{v}}-\frac{3\mu_\mathrm{f}}{H_\mathrm{t}^2}\bar{\bm{v}}
    \label{eq:navie_eq_2d}
  \end{equation}

\subsection{Governing equation of Thermal-fluid layer}

The governing equations of thermal convection in thermal-fluid layer can be expressed as follows:  
\begin{equation}
  \centering
  \rho_\mathrm{f} c_\mathrm{pf} \bm{v}\cdot \nabla T=k\nabla^2T
  \label{eq:thermal-fluid}
\end{equation}
\par
Supposing, axial conduction is negligible \cite{Yanetal}. 
Eq.~\eqref{eq:thermal-fluid} can be approximated as follows: 
\begin{equation}
  \centering
  \rho_\mathrm{f} c_\mathrm{pf} \bm{v} \cdot \nabla T\approx k\frac{\partial^2 T }{\partial z^2}
  \label{eq:thermal-fluid-apporoximate}
\end{equation}
\par
Defining non-dimensional temperature as: 
\begin{equation}
  \centering
  \theta_\mathrm{t}=\frac{T_\mathrm{w}(x,y)-T(x,y,z)}{T_\mathrm{w}(x,y)-T_0(x,y)}
  \label{eq:dimensionless_temp}
\end{equation}
\par
Here, $T_\mathrm{w}$ refers to temperature of bottom surface, i.e., the temperature at the interface between the thermal-fluid and the base-solid layers and $T_0$ refers to bulk temperature of thermal fluid layer. 
Since assuming a completely developed temperature field in constant heat flux, 
the dimensionless temperature distribution is invariant in the axial direction.  
Therefore, the following equations can be satisfied: 
\begin{equation}
  \centering
  \theta_\mathrm{t}=g(z)
  \label{eq:fully_developped_theta}
\end{equation}
\begin{equation}
  \centering
  \frac{\partial T}{\partial z}=-g'(z)(T_\mathrm{w}-T_0)
  \label{eq:T_eq}
\end{equation}
\par
Furthermore, axial gradient of temperature is constant along the height:  
\begin{equation}
  \centering
  \frac{\partial T_\mathrm{w}}{\partial x}=\frac{\partial T_0}{\partial x}=\frac{\partial T}{\partial x}
  \label{eq:fully_developped_T_x}
\end{equation}
\begin{equation}
  \centering
  \frac{\partial T_\mathrm{w}}{\partial y}=\frac{\partial T_0}{\partial y}=\frac{\partial T}{\partial y}
  \label{eq:fully_developped_T_y}
\end{equation}
\par
Substituting Eq.~\eqref{eq:velocity_profile} and \eqref{eq:T_eq} into \eqref{eq:thermal-fluid-apporoximate} gives: 
\begin{equation}
  \centering
  g''(z)=-\lambda(x,y)\left[1-\left(\frac{z}{H_\mathrm{t}}\right)^2\right]
  \label{eq:g_2}
\end{equation}
\begin{equation}
  \centering
  \lambda\equiv-\frac{3}{2} \frac{\rho_\mathrm{f} c_\mathrm{pf}}{k(T_\mathrm{w}-T_0)}  \left(\bar{u}\frac{\partial T_0}{\partial x}+\bar{v}\frac{\partial T_0}{\partial y}\right)
  \label{eq:lambda}
\end{equation} 
\par
Eq.~\eqref{eq:g_post} can be derived by integrating Eq~\eqref{eq:g_2} twice with respect to $z$, subject to the boundary conditions specified in Eqs.~\eqref{eq:bottom_bound_T} and \eqref{eq:upper_bound_T}: 
\begin{equation}
  \centering
  g(z)=\frac{\lambda}{12} \left(13+\frac{8z}{H_\mathrm{t}}-\frac{6z^2}{H_\mathrm{t}^2}+\frac{z^4}{H_\mathrm{t}^4}\right)
  \label{eq:g_post}
\end{equation}
\begin{equation}
  \centering
  T(z=-H_\mathrm{t})=T_\mathrm{w}
  \label{eq:bottom_bound_T}
\end{equation}
\begin{equation}
  \centering
  \frac{\partial T}{\partial z}(z=H_\mathrm{t})=0
  \label{eq:upper_bound_T}
\end{equation}
\par
The definition of the bulk temperature of thermal-fluid can be expressed as: 
\begin{equation}
  \centering
  T_0=\frac{\int_{-H_\mathrm{t}}^{H_\mathrm{t}} \bm{v}T dz}{\int_{-H_\mathrm{t}}^{H_\mathrm{t}} \bm{v} dz}=\frac{1}{2H_\mathrm{t}\bar{v}}\int_{-H_\mathrm{t}}^{H_\mathrm{t}} \bm{v}T dz
  \label{eq:T_bulk}
\end{equation}
\par
Substituting Eq.~\eqref{eq:g_post} into Eq.~\eqref{eq:T_bulk} gives: 
\begin{equation}
  \centering
  g(z)=\frac{T_\mathrm{w}(x, y)-T(x, y, z)}{T_\mathrm{w}(x, y)-T_0(x, y)}=\frac{35}{416} \left(13+\frac{8z}{H_\mathrm{t}}-\frac{6z^2}{H_\mathrm{t}^2}+\frac{z^4}{H_\mathrm{t}^4}\right)
  \label{eq:g}
\end{equation}
\par
The reduced two-dimensional thermal fluid equation can be derived as Eq.~\eqref{eq:thermal_fluid_2d} by substituting Eq.~\eqref{eq:velocity_profile} and \eqref{eq:g} into \eqref{eq:thermal-fluid} as: 
\begin{equation}
  \centering
  \rho_\mathrm{f} c_\mathrm{pf} \bar{v} \cdot \nabla T_0=k\nabla^2 T_0+\frac{h_\mathrm{t}}{2H_\mathrm{t}}(T_\mathrm{w}-T_0)
  \label{eq:thermal_fluid_2d}
\end{equation}
where $h_\mathrm{t}$ is the heat transfer coefficient in thermal-fluid layer expressed as follows: 

\begin{equation}
  \centering
  h_\mathrm{t}=\frac{35k}{26H_\mathrm{t}}
  \label{eq:definition_ht}
\end{equation}

\subsection{Governing equation of Base solid layer}
The governing equation for solid heat conduction can be expressed as follows:
\begin{equation}
  \centering
  k_\mathrm{b} \nabla^2 T_\mathrm{b} =0
  \label{eq:base-solid}
\end{equation}
\par
Boundary condition for isothermal heat flux from the bottom surface can be expressed as follows: 
\begin{equation}
  \centering
  k_\mathrm{b}\frac{\partial T_\mathrm{b}}{\partial z}(z=-H_\mathrm{b})=-q_\mathrm{s}
  \label{eq:bottom_bound_Tb}
\end{equation}


\begin{equation}
  \centering
  k_\mathrm{b}\frac{\partial T_\mathrm{b}}{\partial z}(z=H_\mathrm{b})=-h_\mathrm{b}(T_\mathrm{b0}-T_\mathrm{w})
  \label{eq:upper_bound_Tb}
\end{equation}
where $h_\mathrm{b}$ denotes heat transfer coefficient introduced in analogy with Newton\textquotesingle s~law of cooling, relating $T_\mathrm{b0}$ to $T_\mathrm{w}$ under a uniform heat flux.

Assuming that the thickness of base plate is thin enough, following approximation given in Eq.~\eqref{eq:T_b_apporoximate} can be satisfied:  
\begin{equation}
  \centering
  k_\mathrm{b} \frac{\partial^2 T_\mathrm{b}}{\partial z^2}\approx0
  \label{eq:T_b_apporoximate}
\end{equation}

The temperature profile of the base plate is obtained which is a linear function of $z$ as follows: 

\begin{equation}
  \centering
  T_\mathrm{b}(x,y,z)=T_\mathrm{b0}+\frac{T_w-T_\mathrm{b0}}{H_\mathrm{b}}z
  \label{eq:T_b_distribute}
\end{equation}
\par
Substituting Eq.~\eqref{eq:T_b_distribute} into \eqref{eq:bottom_bound_Tb} gives: 
\begin{equation}
  \centering
  q_\mathrm{s}=k_\mathrm{b}\frac{T_\mathrm{b0}-T_\mathrm{w}}{H_\mathrm{b}}
  \label{eq:qs_T}
\end{equation}
\par
Since the fixed heat flux $q_\mathrm{s}$ is uniform and independent of the $x$--$y$ directions,
following equation can be satisfied: 
\begin{equation}
  \centering
  \frac{\partial T_\mathrm{w}}{\partial x}=\frac{\partial T_\mathrm{b0}}{\partial x}=\frac{\partial T_\mathrm{b}}{\partial x}
  \label{eq:Tb_invariant_x}
\end{equation}

\begin{equation}
  \centering
  \frac{\partial T_\mathrm{w}}{\partial y}=\frac{\partial T_\mathrm{b0}}{\partial y}=\frac{\partial T_\mathrm{b}}{\partial y}
  \label{eq:Tb_invariant_y}
\end{equation}

Substitute Eq.~\eqref{eq:T_b_distribute} into \eqref{eq:upper_bound_Tb}, 
following equation can be derived: 

\begin{equation}
  \centering
  h_\mathrm{b}=\frac{k_\mathrm{b}}{H_\mathrm{b}}
  \label{eq:hb}
\end{equation}

Eq.~\eqref{eq:base-solid_2d_pre} can be derived by averaging Eq.~\eqref{eq:base-solid} over the domain $z\in[-H_\mathrm{t} , H_\mathrm{t}]$: 
\begin{equation}
  \centering
  k_\mathrm{b} \nabla^2 T_\mathrm{b0}+\frac{k_\mathrm{b}}{2H_\mathrm{b}} \left[\frac{\partial T}{\partial z}(z=H_\mathrm{b})-\frac{\partial T}{\partial z}(z=-H_\mathrm{b}) \right]=0
  \label{eq:base-solid_2d_pre}
\end{equation}

Considering boundary conditions in Eq.~\eqref{eq:bottom_bound_Tb} and \eqref{eq:upper_bound_Tb}, 
the reduced two-dimensional governing equation of solid heat conduction can be derived as follows:  

\begin{equation}
  \centering
  k_\mathrm{b} \nabla^2 T_\mathrm{b0}+\frac{q_\mathrm{s}}{2H_\mathrm{b}}-\frac{h_\mathrm{b}}{2H_\mathrm{b}}(T_\mathrm{b0}-T_\mathrm{w})=0
  \label{eq:base-solid_2d}
\end{equation}

\subsection{Coupling of velocity and temperature field in thermal-fluid layer and base solid layer}
The governing equation presented in Subsection 2.2 is derived by coupling Eqs. \eqref{eq:continuity_eq_2d}, \eqref{eq:navie_eq_2d}, \eqref{eq:thermal_fluid_2d}, and \eqref{eq:base-solid_2d}, while incorporating porous approximations such as the DF law and equivalent physical properties.

\section*{CRediT authorship contribution statement}
\noindent \textbf{Tatsuki Saito}: Conceptualization, Methodology,Validation, Visualization, Investigation, Data curation, Formal analysis, Writing - original draft.\\
\textbf{Yuto Kukuchi}: Conceptualization, Methodology, Writing - review \& editing.\\
\textbf{Kuniharu Ushijima}: Supervision, Methodology, Writing - review \& editing, Project administration.\\
\textbf{Kentaro Yaji}: Supervision, Methodology, Writing - review \& editing, Project administration.\\

\section*{Acknowledgements}
This work was partially supported by JSPS KAKENHI Grant Number 23K26018.

\section*{Data availability}
Data is available on request

\section*{Declaration of generative AI and AI-assisted technologies in the writing process.}
During the preparation of this work the authors used such as ChatGPT and DeepL Translate to support translation from Japanese to English and for text correction. 
After using these services, the authors reviewed and edited the content as needed and take full responsibility for the content of the published article.

\section*{Declaration of competing interest}
The authors declare that they have no known competing financial interests or personal relationships that could have appeared to influence the work reported in this paper.







\end{document}